\documentclass[letterpaper,twocolumn,twoside]{IEEEtran} 
\usepackage{amsmath, amssymb, bm, cite, epsfig, psfrag}
\usepackage{algorithm,algorithmic}
\usepackage{bbm}
\usepackage[usenames]{color}
\usepackage{tikz}
\usetikzlibrary{shapes,arrows}
\usetikzlibrary{patterns}
\usepackage{etoolbox}
\usepackage{mathtools}
\usepackage{epstopdf}
\usepackage{url,hyperref}

\newtoggle{conference}
\togglefalse{conference} 

\usepackage[normalem]{ulem} 

\newcommand{\textb}[1]{\textcolor{black}{#1}}
\newcommand{\bluecolor}{\color{black}}
\newcommand{\sz}{0.9} 

\def\beq{\begin{equation}}
\def\eeq{\end{equation}}
\def\beqa{\begin{eqnarray}}
\def\eeqa{\end{eqnarray}}
\def\beqan{\begin{eqnarray*}}
\def\eeqan{\end{eqnarray*}}

\def\R{{\mathbb{R}}}
\def\C{{\mathbb{C}}}
\def\argmin{\mathop{\mathrm{arg\,min}}}
\def\argmax{\mathop{\mathrm{arg\,max}}}
\def\diag{\mathop{\mathrm{Diag}}}
\DeclareMathOperator{\prox}{prox} 

\def\x{\times}

\newcommand*\dif{\mathop{}\!\mathrm{d}} 

\newtheorem{definition}{Definition}
\newtheorem{theorem}{Theorem}
\newtheorem{lemma}{Lemma}

\renewcommand{\tilde}[1]{\widetilde{#1}}

\def\qbar{\overline{q}}

\def\arr{\rightarrow}
\def\Exp{\mathbb{E}}

\def\tm1{t\! - \! 1}
\def\tp1{t\! + \! 1}

\newcommand{\zero}{\mathbf{0}}
\newcommand{\one}{\mathbf{1}}
\newcommand{\abf}{\mathbf{a}}
\newcommand{\bbf}{\mathbf{b}}
\newcommand{\cbf}{\mathbf{c}}

\newcommand{\hbf}{\mathbf{h}}
\newcommand{\pbf}{\mathbf{p}}

\newcommand{\pbfbar}{{\mathbf{p}}}
\newcommand{\qbf}{\mathbf{q}}

\newcommand{\sbf}{\mathbf{s}}
\newcommand{\sbfbar}{{\mathbf{s}}}

\newcommand{\vbf}{\mathbf{v}}

\newcommand{\wbf}{\mathbf{w}}
\newcommand{\rbf}{\mathbf{r}}

\newcommand{\xbf}{\mathbf{x}}
\newcommand{\xbfhat}{\widehat{\mathbf{x}}}
\newcommand{\xbfbar}{\overline{\mathbf{x}}}
\newcommand{\ybf}{\mathbf{y}}
\newcommand{\zbf}{\mathbf{z}}

\newcommand{\Abf}{\mathbf{A}}

\newcommand{\Bbf}{\mathbf{B}}

\newcommand{\Dbf}{\mathbf{D}}
\newcommand{\Fbf}{\mathbf{F}}
\newcommand{\Gbf}{\mathbf{G}}
\newcommand{\Hbf}{\mathbf{H}}
\newcommand{\Ibf}{\mathbf{I}}
\newcommand{\Jbf}{\mathbf{J}}

\newcommand{\Qbf}{\mathbf{Q}}
\newcommand{\Rbf}{\mathbf{R}}

\newcommand{\Sbf}{\mathbf{S}}
\newcommand{\Tbf}{\mathbf{T}}
\newcommand{\Ubf}{\mathbf{U}}
\newcommand{\Vbf}{\mathbf{V}}

\def\taubf{\boldsymbol{\textb{\tau}}}

\def\nubf{\textb{\boldsymbol{\nu}}}

\newcommand{\MAP}{_{\text{\sf MAP}}}
\newcommand{\MMSE}{_{\text{\sf MMSE}}}

\newcommand{\tran}{^{\text{\sf T}}}
\newcommand{\herm}{^{\text{\sf H}}}
\newcommand{\gsbar}{{g}_s}
\newcommand{\pbar}{{p}}
\newcommand{\kappamax}{\kappa_{\max}(\theta_s,\theta_x)}
\newcommand{\kappamaxnodamp}{\kappa_{\max}(1,1)}

\newcommand{\Imag}{\mbox{Imag}}

\tikzstyle{block}=[rectangle,draw, fill=blue!20,
    minimum height=1cm, minimum width=1.5cm]
\tikzstyle{signal}=[coordinate,draw]

\title{On the Convergence of Approximate Message Passing with Arbitrary Matrices}

\iftoggle{conference}{
   \author{
     \IEEEauthorblockN{
        Sundeep Rangan\IEEEauthorrefmark{1},
        Philip Schniter\IEEEauthorrefmark{2}, and
        Alyson K.\ Fletcher\IEEEauthorrefmark{3} \\
     }
     \IEEEauthorblockA{
        \IEEEauthorrefmark{1}NYU, Elec.\ \& Comp.\ Engineering, srangan@poly.edu \\
     }
      \IEEEauthorblockA{
        \IEEEauthorrefmark{2}The Ohio State Univ., Elec.\ \& Comp.\ Engineering, schniter@ece.osu.edu \\
     }
      \IEEEauthorblockA{
        \IEEEauthorrefmark{3}U.C.\ Santa Cruz., Elec.\ Engineering, afletcher@ucsc.edu
     }
   }
}{
  \author{
     Sundeep Rangan, 
     Philip Schniter,  
     Alyson K.\ Fletcher, and
     Subrata Sarkar\\
     \thanks{S. Rangan (email: srangan@nyu.edu) is with
          the Department of Electrical and Computer Engineering,
          New York University, Brooklyn, NY, 11201.
          His work was supported in part by
          the National Science Foundation under Grant 1116589
          and the industrial affiliates of NYU WIRELESS.}
     \thanks{A.~K.~Fletcher (email: akfletcher@ucla.edu) is with
          the Department of Statistics and Electrical Engineering,
          the University of California, Los Angeles, CA, 90095.
          Her work was supported in part by
          the National Science Foundation under Grant 1254204 and
          the Office of Naval Research under Grant N00014-15-1-2677.}
     \thanks{P.~Schniter and S.~Sarkar
          (email: schniter@ece.osu.edu and sarkar.51@osu.edu) are with
          the Department of Electrical and Computer Engineering,
          The Ohio State University, Columbus, OH, 43210.
          Their work was supported in part by
          the National Science Foundation under Grants
          CCF-1018368, CCF-1218754, and CCF-1527162.}
     \thanks{Portions of this work were presented at the 
          IEEE International Symposium on Information Theory
          \cite{RanSchFle:14-ISIT}.}
 }
}

\begin{document}
\setlength{\arraycolsep}{0.8mm}

\maketitle
\begin{abstract}
Approximate message passing (AMP) methods and their variants
have attracted considerable recent
attention for the problem of estimating a random vector $\xbf$
observed through a linear transform $\Abf$.
In the case of large i.i.d.\ zero-mean Gaussian $\Abf$, the methods
exhibit fast convergence with precise analytic characterizations
on the algorithm behavior.
However, the convergence of AMP under general transforms $\Abf$ is not fully understood.
In this paper, we provide sufficient conditions for the convergence
of a damped version of the generalized AMP (GAMP)
algorithm in the case of quadratic cost functions
(i.e., Gaussian likelihood and prior).
It is shown that, with sufficient damping, the algorithm is guaranteed to
converge, although the amount of damping grows with peak-to-average ratio
of the squared singular values of the transforms $\Abf$.  This result explains
the good performance of AMP on i.i.d.\ Gaussian transforms $\Abf$, but
also their difficulties with ill-conditioned or non-zero-mean transforms $\Abf$.
A related sufficient condition is then derived for the local stability
of the damped GAMP method under general cost functions,
assuming certain strict convexity conditions.
\end{abstract}


\section{Introduction}

Consider estimating a random vector $\xbf \in \R^n$ with independent components
$x_j \sim P(x_j)$ from observations $\ybf \in \R^m$ that are conditionally
independent given the transform outputs
\beq \label{eq:GLM}
    \zbf = \Abf\xbf,
\eeq
i.e., $P(\ybf|\zbf)=\prod_i P(y_i|z_i)$.
Here, we assume knowledge of the matrix $\Abf \in \R^{m \x n}$ in \eqref{eq:GLM}
and the densities $P(x_j)$ and $P(y_i|z_i)$.
Often, the goal is to compute either the minimum mean-squared error (MMSE) estimate
$\xbfhat\MMSE = \int_{\R^n} \!\xbf\, P(\xbf|\ybf) \dif\xbf = \Exp(\xbf|\ybf)$
or the maximum a posteriori (MAP) estimate
$\xbfhat\MAP = \argmax_{\xbf \in \R^n} P(\xbf|\ybf)$,
where in either case $P(\xbf|\ybf)$ denotes the posterior distribution.
Using $F(\zbf):=-\ln P(\ybf|\zbf)$ and $G(\xbf):=-\ln P(\xbf)$
and Bayes rule, $P(\xbf|\ybf)\propto P(\ybf|\xbf)P(\xbf)$,
it becomes evident that MAP estimation is equivalent to the optimization problem
\beqa
    \xbfhat\MAP
    = \argmin_{\xbf \in \R^n} F(\Abf\xbf) + G(\xbf)
	\label{eq:MAP}
\eeqa
for separable $F(\zbf)=\sum_i F_i(z_i)$ and $G(\xbf)=\sum_j G_j(x_j)$.
Such problems arise in a range of applications including
statistical regression,
inverse problems, and
compressed sensing.

Most current numerical methods for solving the constrained
optimization problem \eqref{eq:MAP}
attempt to exploit the separable
structure of the objective function \eqref{eq:MAP}
using approaches like iterative shrinkage and thresholding (ISTA)
\cite{ChamDLL:98,DaubechiesDM:04,WrightNF:09,BeckTeb:09,Nesterov:07,BioDFig:07},
the alternating direction method of multipliers (ADMM)
\cite{BoydPCPE:09,Esser:JIS:10,Chambolle:JMIV:11,He:JIS:12},
or primal-dual approaches
\cite{Esser:JIS:10,Chambolle:JMIV:11,He:JIS:12,Komodakis:SPM:15}.

In recent years, however, there has also been considerable
interest in approximate message passing (AMP) methods
that apply Gaussian and quadratic approximations to loopy
belief propagation (BP) in graphical models
\cite{DonohoMM:09,DonohoMM:10-ITW1,Rangan:11-ISIT}.
AMP applied to max-sum loopy BP produces a sequence of estimates that approximate $\xbfhat\MAP$, while AMP applied to sum-product loopy BP produces a sequence of estimates that approximate $\xbfhat\MMSE$.
For zero-mean i.i.d.\ sub-Gaussian $\Abf$
in the large-system limit (i.e., $m,n\rightarrow\infty$ with fixed $m/n$),
AMP methods are characterized by a state evolution
whose fixed points, when unique, coincide with $\xbfhat\MAP$ or $\xbfhat\MMSE$
\cite{BayatiM:11,javanmard2013state,BayLelMon:15}.
In addition, for large but finite-sized i.i.d.\ Gaussian matrices, recent work \cite{rush2016finite} shows that AMP is close to Bayes-optimal.

Unfortunately, a rigorous characterization of AMP for generic $\Abf$ remains lacking.
The recent papers \cite{RanSRFC:13-ISIT,Krzakala:14-ISITbethe} studied the fixed-points of the generalized
AMP (GAMP) algorithm from \cite{Rangan:11-ISIT} for generic $\Abf$.
In \cite{RanSRFC:13-ISIT}, it was established that the fixed points of max-sum GAMP coincide with the critical points of the optimization objective in \eqref{eq:MAP}.
Similarly, \cite{RanSRFC:13-ISIT,Krzakala:14-ISITbethe} established that
the fixed points of sum-product GAMP are critical points of a large-system version of the Bethe free energy from \cite{YedidiaFW:03}.
However, the papers \cite{RanSRFC:13-ISIT,Krzakala:14-ISITbethe} did not discuss the convergence of the algorithm
to those fixed points.
Indeed, similar to other loopy BP algorithms, GAMP may diverge, as demonstrated for mildly ill-conditioned $\Abf$ in \cite{Vila:ICASSP:15}.
Likewise, \cite{Caltagirone:14-ISIT}
showed that AMP can diverge with non-zero-mean i.i.d.\ Gaussian $\Abf$
and the divergence can, in fact, be predicted via a
state-evolution analysis.

For general loopy BP, a variety of methods
have been proposed to improve convergence, including coordinate descent,
tree re-weighting, and double loop methods
\cite{pretti2005message,kolmogorov2006convergent,globerson2007fixing,manoel2015swamp,rangan2015admm}.
In this paper, we propose and analyze a ``damped'' modification of GAMP that is similar
to the technique used in Gaussian belief propagation \cite{bickson2008gaussian,dolev2009fixing}---a closely related algorithm.
We also point out connections between damped GAMP and the primal-dual hybrid-gradient (PDHG) algorithm
\cite{Esser:JIS:10,Chambolle:JMIV:11,He:JIS:12,Komodakis:SPM:15}
popular in convex optimization.
This connection enhances the interpretability of AMP methods, especially for those who are less familiar with belief propagation.

Our first main result
establishes a necessary and sufficient
condition on the global convergence of damped GAMP
for arbitrary $\Abf$ in the special case of Gaussian $P(x_j)$ and $P(y_i|z_i)$
(i.e., quadratic $F$ and $G$) and fixed scalar stepsizes.
This condition (see Theorem~\ref{thm:linConvUnif} below) shows that, with sufficient
damping, the Gaussian GAMP algorithm can be guaranteed to converge.
However, the amount of damping grows with the peak-to-average ratio of the
squared singular values of $\Abf$.
This result explains why Gaussian GAMP converges (with high probability)
for large i.i.d.\ Gaussian $\Abf$, but it also explains why it needs
to be damped significantly for non-zero-mean, low-rank,
or otherwise ill-conditioned $\Abf$.

Our second result establishes the local convergence of GAMP for strictly convex
$F$ and $G$ and arbitrary, but fixed, vector-valued stepsizes.
This sufficient condition is similar to the Gaussian case, but involves a certain
row-column normalized version of $\Abf$.
(See Theorem~\ref{thm:localStab} below.)
\iftoggle{conference}{
For space considerations, proofs are omitted but can be found in a full paper
\cite{RanSchFle:14}, which also contains more discussion and numerical experiments.}
{}

Finally, we present numerical experiments that verify the tightness of the sufficient conditions from Theorems~\ref{thm:linConvUnif} and~\ref{thm:localStab}.

\emph{Notation}:
We use
capital boldface letters like $\Abf$ for matrices,
small boldface letters like $\abf$ for vectors,
$(\cdot)\tran$ for transposition,
$(\cdot)\herm$ for Hermitian (i.e., conjugate transposition), and
$a_i=[\abf]_i$ to denote the $i$th element of $\abf$.
Also, we use
$\|\Abf\|_2$ for the spectral norm of $\Abf$,
$\|\Abf\|_F$ for the Frobenius norm of $\Abf$,
and
$\diag(\abf)$ for the diagonal matrix created from vector $\abf$.
In addition, we use
$\zero$ for the all-zeros vector,
$\one$ for the all-ones vector,
and
$\Ibf_N$ for the $N\times N$ identity matrix.
Although it is somewhat non-standard, we use
$\Abf.\Bbf$ for component-wise multiplication,
$\Abf./\Bbf$ for component-wise division of the matrices $\Abf$ and $\Bbf$,
and $|\Abf|$ for component-wise magnitude of $\Abf$.
\textb{Similarly, we use $\abf\geq\zero$ to denote component-wise inequality (i.e., $a_i\geq 0$ for $i=1,..,n$).}
For a random vector $\xbf$, we denote
its probability density function (pdf) by $P(\xbf)$,
and
its expectation by $\Exp[\xbf]$.
Similarly, we use
$P(\xbf|\ybf)$
and
$\Exp[\xbf|\ybf]$
for the \emph{conditional}
pdf
and
expectation,
respectively.
We refer to
the pdf of a Gaussian random vector $\xbf\in\R^N$ with mean $\abf$ and covariance $\Rbf$ using ${\mathcal N}(\xbf;\abf,\Rbf)=\exp( -(\xbf-\abf)\tran\Rbf^{-1}(\xbf-\abf)/2 )/\sqrt{(2\pi)^N|\Rbf|}$.
Finally,
$P(\xbf)\propto Q(\xbf)$ says that functions $P(\cdot)$ and $Q(\cdot)$ are equal up to a scaling that is invariant to $\xbf$.

\section{Damped GAMP}

\subsection{Review of GAMP}
The GAMP algorithm was introduced in \cite{Rangan:11-ISIT}
and rigorously analyzed in \cite{javanmard2013state}.
The procedure (see Algorithm~\ref{algo:gamp}) produces a sequence of estimates $\xbfhat^t, t=1,2,\dots$, that, in max-sum mode, approximate $\xbfhat\MAP$ and, in sum-product mode, approximate $\xbfhat\MMSE$.
The two modes differ only in the definition of the scalar estimation functions $\gsbar$ and $g_x$
used in lines~\ref{line:taust}, \ref{line:st}, \ref{line:tauxt}, and \ref{line:xt} of Algorithm~\ref{algo:gamp}:
\begin{itemize}
\item
In \emph{max-sum} mode,
\beqa
    {[g_x(\rbf,\taubf_r)]}_j
    &=& \prox_{\tau_{r_j}G_j}(r_j)
    	\label{eq:gxMAP} \\
    {[\gsbar(\pbfbar,\nubf_p)]}_i
    &=& \pbar_i - \nu_{p_i} \prox_{F_i/\nu_{p_i}}(\pbar_i/\nu_{p_i})
    	\label{eq:gsbarMAP}
\eeqa
using $\taubf_r=[\tau_{r_1},\dots,\tau_{r_n}]\tran$,
$\nubf_p=[\nu_{p_1},\dots,\nu_{p_m}]\tran$,
and
\beq
    \prox_{f}(r)
    := \argmin_{x} f(x) + \tfrac{1}{2}|x-r|^2 .
    	\label{eq:prox}
\eeq
Note \eqref{eq:gxMAP} implements scalar MAP denoising under prior $P(x_j)\!\propto\! \exp(-G(x_j))$ and
variance-$\tau_{r_j}$ Gaussian noise.

\item
In \emph{sum-product} mode,
\beqa
    {[g_x(\rbf,\taubf_r)]}_j
    &=& \frac{
      \int x_j P(x_j)
      \mathcal{N}(x_j;r_j,\tau_{r_j})
      \dif x_j}
      {\int P(x_j)
      \mathcal{N}(x_j;r_j,\tau_{r_j})
      \dif x_j}
    	\label{eq:gxMMSE}\\
    {[\gsbar(\pbfbar,\nubf_p)]}_i
    &=& \pbar_i - \nu_{p_i}
      \frac{
      \int z_i P(y_i|z_i)
      \mathcal{N}(z_i;\frac{\pbar_i}{\nu_{p_i}},\frac{1}{\nu_{p_i}})
      \dif z_i}
      {\int P(y_i|z_i)
      \mathcal{N}(z_i;\frac{\pbar_i}{\nu_{p_i}},\frac{1}{\nu_{p_i}})
      \dif z_i} ,
      \nonumber
\eeqa
and so \eqref{eq:gxMMSE} is the scalar MMSE denoiser under $P(x_j)\!\propto\! \exp(-G(x_j))$ and variance-$\tau_{r_j}$ Gaussian noise.
\end{itemize}
Note that, in Algorithm~\ref{algo:gamp} and the sequel, $\abf.\bbf$ and $\abf./\bbf$ denote component-wise multiplication and division, respectively, between vectors $\abf$ and $\bbf$.

\begin{algorithm}[t]
\caption{GAMP with vector stepsizes and damping}
\begin{algorithmic}[1]  \label{algo:gamp}
\REQUIRE{
Matrix $\Abf$,
scalar estimation functions $g_x$ and $\gsbar$,
and damping constants $\theta_s, \theta_x \in (0,1]$. }

\STATE{ $\Sbf = \Abf.\Abf$ (component-wise magnitude squared)} \label{line:Sdef}
\STATE{ $t = 0$  }
\STATE{ Initialize $\taubf_x^t>\zero$, $\xbf^t$  }
\STATE{ $\sbfbar^{\tm1} = \zero$ }
\REPEAT
    \STATE{ $\one./\nubf_p^t = \Sbf\taubf_x^t$ }\label{line:taupt}
    \STATE{ $\pbfbar^t = \sbfbar^{\tm1} + \nubf_p^t.\Abf\xbf^t$ } \label{line:pt}
    \STATE{ $\nubf_s^t = \nubf_p^t.\gsbar'(\pbfbar^t,\nubf_p^t)$ } \label{line:taust}
    \STATE{ $\sbfbar^t = (1-\theta_s)\sbfbar^{\tm1} + \theta_s\gsbar(\pbfbar^t,\nubf^t_p)$ } \label{line:st}
    \STATE{ $\one./\taubf_r^t = \Sbf\tran\nubf_s^t$ } \label{line:taurt}
    \STATE{ $\rbf^t = \xbf^t - \taubf^t_r .\Abf\herm\sbfbar^t$ } \label{line:rt}
    \STATE{ $\taubf^{\tp1}_x = \taubf^t_r.g_x'(\rbf^t, \taubf^t_r )$ }
        \label{line:tauxt}
    \STATE{ $\xbf^{\tp1} = (1-\theta_x)\xbf^t + \theta_x g_x(\rbf^t, \taubf^t_r)$ }
	\label{line:xt}
    \STATE{ $t\gets t+1$ }
\UNTIL{Terminated}

\end{algorithmic}
\end{algorithm}

Algorithm~\ref{algo:gamp} reveals the computational efficiency of GAMP:
the vector-valued MAP and MMSE estimation problems are reduced to
a sequence of scalar estimation problems in Gaussian noise.
Specifically, each iteration involves multiplications by $\Sbf$,
$\Sbf\tran$, $\Abf$ and $\Abf\herm$
along with simple scalar estimations on the components $x_j$ and $z_i$;
there are no vector-valued estimations or matrix inverses.

We note that Algorithm~\ref{algo:gamp} writes GAMP in a ``symmetrized'' form,
where the steps in lines \ref{line:taupt}-\ref{line:st} mirror those
in lines \ref{line:taurt}-\ref{line:xt}.
\textb{This differs from the way that GAMP is presented in most other publications, such as \cite{Rangan:11-ISIT}, which is obtained by replacing the variables $\sbf$, $\nubf_p$, and $\pbf$ in Algorithm~\ref{algo:gamp} by $-\sbf$, $\one./\taubf_p$, and $\pbf.\taubf_p$, respectively.
Note that, thoughout this paper, we use $\tau$ for variance quantities and $\nu$ for precision (i.e., inverse variance) quantities. }

\subsection{Damped GAMP} \label{sec:damp}

Algorithm~\ref{algo:gamp} includes a small but important modification to the
original GAMP from \cite{Rangan:11-ISIT}:
lines \ref{line:st} and \ref{line:xt} perform \emph{damping} using constants
$\theta_s,\theta_x \in (0,1]$ that slow the updates of $\sbf^t,\xbf^t$
when $\theta_s, \theta_x < 1$, respectively.
The original GAMP implicitly uses $\theta_s=1=\theta_x$.
In the sequel, we establish---analytically---that damping facilitates the convergence of GAMP for general $\Abf$, a fact that has been empirically observed in past works
(e.g., \cite{Schniter:ALL:12,Caltagirone:14-ISIT,Vila:ICASSP:15}).

\subsection{GAMP with Scalar Stepsizes} \label{sec:gampUnif}

The computational complexity of Algorithm~\ref{algo:gamp} is dominated by the
matrix-vector multiplications involving $\Abf$, $\Abf\herm$, $\Sbf,$ and $\Sbf\tran$.
In \cite{Rangan:10arXiv-GAMP}, a \emph{scalar-stepsize} simplification of GAMP
was proposed to avoid the multiplications by $\Sbf$ and $\Sbf\tran$, roughly
halving the per-iteration complexity.
The meaning of ``stepsize'' will become clear in the sequel.
Algorithm~\ref{algo:gampUnif} shows the scalar-stepsize version of
Algorithm~\ref{algo:gamp}.

\iftoggle{conference}{In \cite{RanSchFle:14}, we}{For use in the sequel, we now}
show that scalar-stepsize GAMP is equivalent to
vector-stepsize GAMP under a different choice of $\Sbf$.
While Algorithm~\ref{algo:gamp} uses $\Sbf = \Abf.\Abf$,
Algorithm~\ref{algo:gampUnif} effectively uses
\beq \label{eq:Sunif}
    \Sbf = \frac{\|\Abf\|_F^2}{mn}\one\one\tran ,
\eeq
i.e., a constant matrix having the same average value as $\Abf.\Abf$.
\iftoggle{conference}{}{
Thus, the two algorithms coincide when $|A_{ij}|$ is invariant to $i$ and $j$.
To see the equivalence, we first note that, under $\Sbf$ from \eqref{eq:Sunif},
line~\ref{line:taupt} in Algorithm~\ref{algo:gamp} would produce a version of
$1/\nubf_p^t$ containing identical elements $1/\nu_p^t$, where
\[
    \frac{1}{\nu_p^t}
    = \frac{\|\Abf\|_F^2}{mn}\one\tran\taubf_x^t
    = \frac{\|\Abf\|_F^2}{m} \tau_x^t
\]
for $\tau_x^t = (1/n)\one\tran\taubf_x$.
Similarly, line~\ref{line:taurt} would produce a vector $1/\taubf_r^t$
with identical elements $1/\tau_r^t$, where
\[
    \frac{1}{\tau_r^t}
    = \frac{\|\Abf\|_F^2}{mn} \one\tran\nubf_s^t
    = \frac{\|\Abf\|_F^2}{n} \nu_s^t,
\]
for $\nu_s^t = (1/m)\one\tran\nubf_s^t$.
Furthermore, $\nubf_p^t=\nu_p^t\one$ and line~\ref{line:taust} imply
that $\nu_s^t = (\nu_p^t/m)\one\tran\gsbar(\pbfbar^t,\nubf_p^t)$,
while $\taubf_r^t=\tau_r^t\one$ and line~\ref{line:tauxt} imply
that $\tau_x^{\tp1} = (\tau_r^t/n)\one\tran g_x(\rbf^t,\taubf_r^t)$.
Applying these modifications to Algorithm~\ref{algo:gamp},
we arrive at Algorithm~\ref{algo:gampUnif}.
}

\begin{algorithm}[t]
\caption{GAMP with scalar stepsizes and damping}
\begin{algorithmic}[1]  \label{algo:gampUnif}
\REQUIRE{ Matrix $\Abf$,
scalar estimation functions $g_x$ and
$\gsbar$,
and damping constants $\theta_s, \theta_x \in (0,1]$. }
\STATE{ $t = 0$  }
\STATE{ Initialize $\tau_x^t>0$, $\xbf^t$  }
\STATE{ $\sbfbar^{\tm1} = \zero$ }
\REPEAT
    \STATE{ $1/\nu_p^t = (1/m)\|\Abf\|_F^2\tau_x^t $ }\label{line:tauptunif}
    \STATE{ $\pbfbar^t = \sbfbar^{\tm1} + \nu_p^t \Abf\xbf^t$ } \label{line:ptunif}
    \STATE{ $\nu_s^t = (\nu_p^t/m)\one\tran \gsbar'(\pbfbar^t,\nu_p^t)$ }
        \label{line:taustunif}
    \STATE{ $\sbfbar^t = (1-\theta_s)\sbfbar^{\tm1} + \theta_s
        \gsbar(\pbfbar^t,\nu_p^t)$ } \label{line:stunif}
    \STATE{ $1/\tau_r^t = (1/n)\|\Abf\|^2_F\nu_s^t$ } \label{line:taurtunif}
    \STATE{ $\rbf^t = \xbf^t - \tau^t_r \Abf\herm\sbfbar^t$ } \label{line:rtunif}
    \STATE{ $\tau^{\tp1}_x = (\tau^t_r/n) \one\tran g_x'(\rbf^t, \tau^t_r)$ }
        \label{line:tauxtunif}
    \STATE{ $\xbf^{\tp1} = (1-\theta_x)\xbf^t + \theta_x
        g_x(\rbf^t, \tau^t_r)$ } \label{line:xtunif}
    \STATE{ $t\gets t+1$ }
\UNTIL{Terminated}

\end{algorithmic}
\end{algorithm}

\subsection{Relation to Primal-Dual Hybrid Gradient Algorithms}
An important case of \eqref{eq:MAP} is when $F$ and $G$ are closed
proper convex functionals and the solution $\xbfhat\MAP$ exists.
Recently, there has been great interest in solving this problem from the
\emph{primal-dual} perspective \cite{Esser:JIS:10,Komodakis:SPM:15}, which can be described as follows.
Consider $F^*$, the \emph{convex conjugate} of $F$, as given by the
Legendre-Fenchel transform
\beq
  F^*(\sbf) := \sup_{\zbf\in\R^m} \sbf\tran\zbf - F(\zbf) .
\eeq
For closed proper convex $F$, we have $F^{**}=F$, and so
\beq
  F(\Abf\xbf) = \sup_{\sbf\in\R^m} \sbf\tran\Abf\xbf - F^*(\sbf) ,
\eeq
which gives the equivalent \emph{saddle-point} formulation of \eqref{eq:MAP},
\beq
  \min_{\xbf\in\R^n} \sup_{\sbf\in\R^m}
  \sbf\tran\Abf\xbf - F^*(\sbf) + G(\xbf) .
    \label{eq:saddle}
\eeq
The so-called \emph{primal-dual hybrid-gradient} (PDHG) algorithm recently
studied in \cite{Esser:JIS:10,Chambolle:JMIV:11,He:JIS:12,Komodakis:SPM:15}
is defined by the iteration
\beqa
  \sbfbar^t
  &\gets& \prox_{\nu_p F^*}\big(\sbfbar^{t-1} + \nu_p \Abf\xbf^t\big)
  	\label{eq:PDHGst}\\
  \xbfhat^{\tp1}
  &\gets& \prox_{\tau_r G}\big(\xbfhat^t - \tau_r \Abf\herm\sbfbar^t\big)
  	\label{eq:PDHGxhatt}\\
  \xbf^{\tp1}
  &\gets& \xbfhat^{\tp1} + \theta (\xbfhat^{\tp1}-\xbfhat^t),
  	\label{eq:PDHGxt}
\eeqa
where $\theta\in[-1,1]$ is a relaxation parameter.
Line \eqref{eq:PDHGst} can be recognized as proximal gradient ascent
in the dual variable $\sbfbar$ using stepsize $\nu_p$,
while line \eqref{eq:PDHGxhatt} is proximal gradient descent
in the primal variable $\xbf$ using stepsize $\tau_r$.

PDHG can be related to damped scalar-stepsize GAMP as follows.
Since $F$ is proper, closed, and convex, we can apply
the Moreau identity \cite{Combettes:MMS:05}
\beq
  \pbfbar
  = \prox_{\nubf_p F^*}(\pbfbar) + \nubf_p \prox_{F/\nubf_p}(\pbfbar/\nubf_p)
  \label{eq:Moreau}
\eeq
to \eqref{eq:gsbarMAP}, after which the assumed separability of $F$ implies that
\beq
  [\gsbar(\pbfbar,\nubf_p)]_i
  = \prox_{\nu_{p_i} F_i^*}(\pbar_i) .
  \label{eq:gsbarMAPdual}
\eeq
Thus, under $\theta_s=1$, scalar GAMP's update of $\sbfbar$
(in line~\ref{line:stunif} of Algorithm~\ref{algo:gampUnif})
matches PDHG's in \eqref{eq:PDHGst}.
Similarly, noting the connection between \eqref{eq:gxMAP} and \eqref{eq:PDHGxhatt},
it follows that, under $\theta_x=1$, scalar GAMP's update of $\xbf$
(in line~\ref{line:xtunif} of Algorithm~\ref{algo:gampUnif}))
matches the PDHG update \eqref{eq:PDHGxt} under $\theta=0$.

In summary, PDHG under
$\theta=0$ (the Arrow-Hurwicz \cite{Arrow:Book:58} case)
would be equivalent to non-damped scalar GAMP if the stepsizes
$\nu_p^t$ and $\tau_r^t$ were fixed over the iterations.
GAMP, however, \emph{adapts} these stepsizes.
In fact, under the existence of the second derivative $f''$, it can be shown that
\beq
    \prox'_{f}(r)
    = \big[ 1+ f''\big(\prox_{f}(r)\big) \big]^{-1} ,
    	\label{eq:prox'}
\eeq
implying that, for smooth $F$ and $G$, GAMP updates $\tau_x^t$ according to
the average local curvature of $G$ at the point $\xbf=\prox_{\tau_r^t G}(\rbf^t)$
and updates $\nu_s^t$ according to
the average local curvature of $F^*$ at the point
$\sbfbar=\prox_{\nu_p^t F^*}(\pbfbar^t)$.
A different form of PDHG stepsize adaptation has been recently considered in
\cite{Goldstein:13}, one that is not curvature based.

\iftoggle{conference}{}{
Meanwhile, PDHG under $\theta\neq 0$ is similar to fixed-stepsize damped scalar GAMP
with $\theta_s=1$ and $\theta_x=1+\theta$, although not the same.
Note that PDHG uses the damped version of $\xbf$ only in the dual update \eqref{eq:PDHGst}
whereas GAMP uses the damped version of $\xbf$ in both primal and dual updates.
Also, PDHG relaxes only the primal variable $\xbf$, whereas damped GAMP relaxes (or damps) both
primal and dual variables.
}

\section{Damped Gaussian GAMP}

\subsection{Gaussian GAMP} \label{sec:gampAwgn}

Although Algorithms~\ref{algo:gamp} and \ref{algo:gampUnif} apply to generic
distributions $P(x_j)$ and $P(y_i|z_i)$, we find it useful to at first
consider the simple case of Gaussian distributions, and in particular
\[
    P(x_j) = {\mathcal N}(x_j;x_{0_j},\tau_{0_j}), \quad
    P(y_i|z_i) =  {\mathcal N}(z_i;y_i,\nu_{w_i}^{-1}),
\]
where $\tau_{0_j}$ are variances and $\nu_{w_i}$ are precisions \textb{(i.e., inverse variances)}.
In this case, the scalar estimation functions used in max-sum mode
are identical to those in sum-product mode, and are linear
\cite{Rangan:10arXiv-GAMP}:
\begin{subequations} \label{eq:gsxLin}
\beqa
    \gsbar(\pbfbar,\nubf_p)
    &=& \nubf_w.(\pbfbar + \nubf_w. \ybf)./(\nubf_p + \nubf_w) - \nubf_w. \ybf \qquad \\
    g_x(\rbf,\taubf_r)
    &=& \taubf_0.(\rbf-\xbf_0)./(\taubf_0+\taubf_r) +\xbf_0.
\eeqa
\end{subequations}
Henceforth, we use ``Gaussian GAMP'' (GGAMP) when referring to
GAMP under the estimation functions \eqref{eq:gsxLin}.

\subsection{Convergence of GGAMP Stepsizes}

We first establish the convergence of the GGAMP stepsizes in the case of
an arbitrary matrix $\Abf$.
For the vector-stepsize case in Algorithm~\ref{algo:gamp},
\iftoggle{conference}{the stepsizes follow the recursions
\beq \label{eq:tausxFix}
    \frac{1}{\nubf_s^t}
    = \Sbf\taubf_x^t + \frac{1}{\nubf_w}, \quad
    \frac{1}{\taubf_x^{\tp1}}
    = \Sbf\tran\nubf_s^t + \frac{1}{\taubf_0},
\eeq }{
lines~\ref{line:taust}~and~\ref{line:tauxt} become
\begin{subequations} \label{eq:tausxGen}
\beqa
    \nubf_s^t
    &=& \nubf_p^t .\gsbar'(\pbfbar^t,\nubf^t_p)
    = \nubf_p^t.\nubf_w./(\nubf_p^t+\nubf_w) \\
    \taubf_x^{\tp1}
    &=& \taubf_r^t.g_x'(\rbf^t,\taubf_r^t)
    = \taubf_r^t.\taubf_0./(\taubf_r^t+\taubf_0),
\eeqa
\end{subequations}
and, combining these with lines~\ref{line:taupt}~and~\ref{line:taurt}, we get
\begin{subequations} \label{eq:tausxFix}
\beqa
    1./\nubf_s^t
    &=& \Sbf\taubf_x^t + 1./\nubf_w \label{eq:tausFix} \\
    1./\taubf_x^{\tp1}
    &=& \Sbf\tran\nubf_s^t + 1./\taubf_0, \label{eq:tauxFix}
\eeqa
\end{subequations} }
which are invariant to $\theta_s, \theta_x, \sbfbar^t$, and $\xbf^t$.
The scalar-stepsize case in Algorithm~\ref{algo:gampUnif} is similar, and in
either case, the following
theorem shows that the GGAMP stepsizes always converge.

\medskip
\begin{theorem} \label{thm:varConv}
Consider Algorithms~\ref{algo:gamp} or \ref{algo:gampUnif})
with Gaussian estimation functions \eqref{eq:gsxLin}
defined for any vectors $\nubf_w$ and $\taubf_0 > \bm{0}$.
Then, as $t\rightarrow\infty$,
the stepsizes $\nubf_p^t, \nubf_s^t, \taubf_r^t, \taubf_x^t$
(or their scalar versions)
converge to unique fixed points that are invariant to $\theta_s$ and $\theta_x$.
\end{theorem}
\iftoggle{conference}{}{
\begin{IEEEproof}  See Appendix~\ref{sec:varConvPf}.
\end{IEEEproof}
}

\section{Scalar-Stepsize GGAMP Convergence}

\subsection{Scalar-stepsize GGAMP}

\iftoggle{conference}{}{
\medskip
\textb{
An important special case that we now consider is scalar-stepsize GGAMP
from Algorithm~\ref{algo:gampUnif} under identical variances, i.e.,
\beq \label{eq:constVar}
    \nubf_w = \nu_w\one, \quad \taubf_0 = \tau_0\one,
\eeq
for some $\nu_w$ and $\tau_0 > 0$.
In this case, lines~\ref{line:taustunif}~and~\ref{line:tauxtunif} give
\begin{subequations} \label{eq:gderivLinUnif}
\beqa
    \nu_s^t
    &=& \frac{1}{m} \one\tran \big(\nubf_p^t. \gsbar'(\pbfbar^t,\nubf_p^t)\big)
    = \frac{\nu_p^t\nu_w}{\nu_p^t+\nu_w} \\
    \tau_x^{\tp1}
    &=& \frac{1}{n} \one\tran \big(\taubf_r^t. g_x'(\rbf^t,\taubf_r^t)\big)
    = \frac{\tau_r^t\tau_0}{\tau_r^t+\tau_0},
\eeqa
\end{subequations}
and, combining these with lines~\ref{line:tauptunif}~and~\ref{line:taurtunif}, we get
\begin{subequations} \label{eq:tauFixUnif}
\beqa
    \frac{1}{\nu_s^t} &=& \frac{1}{\nu_p^t} + \frac{1}{\nu_w} =
        \frac{1}{m}\|\Abf\|_F^2\tau_x^t + \frac{1}{\nu_w} \label{eq:tausFixUnif} \\
    \frac{1}{\tau_x^{\tp1}} &=& \frac{1}{\tau_r^t} + \frac{1}{\tau_0}
        = \frac{1}{n}\|\Abf\|_F^2\nu_s^t + \frac{1}{\tau_0}.
            \label{eq:tauxFixUnif}
\eeqa
\end{subequations}
}
}

\subsection{Convergence}

We now investigate the convergence of the primal and dual variables
$\xbf^t$ and $\sbfbar^t$ for scalar GGAMP.
Since, for this algorithm, the previous section established that, as
$t\rightarrow\infty$, the stepsizes $\nu_p^t$ and $\tau_r^t$ converge
independently of $\theta_s,\theta_x,\sbfbar^t$, and $\xbfbar^t$,
we henceforth consider GGAMP with \emph{fixed} stepsizes
$\nu_p^t=\nu_p$ and $\tau_r^t=\tau_r$,
where $\nu_p$ and $\tau_r$ are the fixed points of
\iftoggle{conference}
{\eqref{eq:tausxFix} under Algorithm~\ref{algo:gampUnif}'s
effective definition of $\Sbf$ in \eqref{eq:Sunif}.}
{\eqref{eq:tauFixUnif} for Algorithm~\ref{algo:gampUnif}.}
(A generalization to arbitrary fixed stepsizes will be given in Section~\ref{sec:localStab}.)

\begin{theorem} \label{thm:linConvUnif}
Define
\begin{equation} \label{eq:gamUnif}
    \Gamma(\theta_s,\theta_x) :=
    \begin{cases}
      \displaystyle
      \frac{2\left[(2-\theta_s)m + \theta_sn\right]}{\theta_s\theta_x mn} &
      \text{if } m \geq n \\[3mm]
      \displaystyle
      \frac{2\left[(2-\theta_x)n + \theta_xm\right]}{\theta_s\theta_x mn} &
      \text{if } m \leq n.
    \end{cases}
\end{equation}
Under Gaussian priors (i.e., \eqref{eq:gsxLin}) with identical variances
\iftoggle{conference}{($\nu_w = \nu_w\one$ and
 $\taubf_0 = \tau_0\one$ for scalars $\nu_w$ and $\tau_0$),}
{\eqref{eq:constVar},}
scalar-stepsize GAMP from Algorithm~\ref{algo:gampUnif}
converges for any $\nu_w$ and $\tau_0>0$ when
\beq\label{eq:sigBndUnifSuff}
    \Gamma(\theta_s,\theta_x) > \|\Abf\|_2^2/\|\Abf\|_F^2.
\eeq
Conversely, it diverges for large enough $\tau_0\nu_w$ when
\beq
    \Gamma(\theta_s,\theta_x) < \|\Abf\|_2^2/\|\Abf\|_F^2.
\eeq
\end{theorem}
\iftoggle{conference}{}{
\begin{IEEEproof} See Appendix~\ref{sec:linConvUnifPf}.
\end{IEEEproof}
}
\medskip
Theorem~\ref{thm:linConvUnif} provides a simple necessary and sufficient condition
on the convergence of scalar GGAMP.
To better interpret this condition, recall that
$\|\Abf\|_2^2$ is the maximum squared singular value of $\Abf$
and that $\|\Abf\|_F^2$ is the sum of the squared singular values of $\Abf$
(i.e., $\|\Abf\|^2_F = \sum_{i=1}^{\min\{m,n\}} \sigma^2_i(\Abf)$).
Thus
\beq \label{eq:kappa}
    \kappa(\Abf) := \frac{\|\Abf\|_2^2}{\|\Abf\|^2_F/
        \min\{m,n\} }
\eeq
is the peak-to-average ratio of the squared singular values of $\Abf$.
Convergence condition \eqref{eq:sigBndUnifSuff} can then be rewritten as
\beq \label{eq:kappaCond}
    \kappa(\Abf) < \kappamax
    := \min\{m,n\}\Gamma(\theta_s,\theta_x),
\eeq
meaning that, for GGAMP convergence, it is necessary and sufficient to choose
$\kappamax$ above the
peak-to-average ratio of the squared singular values.

When there is no damping (i.e., $\theta_s=1=\theta_x$),
the definitions in \eqref{eq:gamUnif} and \eqref{eq:kappaCond}
can be combined to yield
\beq \label{eq:kappaNoDamp}
    \kappamaxnodamp
    = \frac{ 2\min\{m,n\} (m+n)}{mn} \in (2,4].
\eeq
More generally, for $\theta_s,\theta_x \in (0,1]$, it can be shown that
\beq \label{eq:kappaBnd}
    \frac{2}{\theta_s\theta_x}
        < \kappamax \leq
    \frac{4}{\theta_s\theta_x},
\eeq
\iftoggle{conference}{and thus GGAMP can be made to converge
by selecting $\theta_s$, $\theta_x$ sufficiently small.}{
so that the necessary and sufficient GGAMP convergence condition
\eqref{eq:kappaCond} can be rewritten as
\beq \label{eq:dampBnd}
    \theta_s\theta_x < \frac{C}{\kappa(\Abf)} \text{ for some $C \in (2,4]$},
\eeq
which implies that, by choosing sufficiently small damping constants
$\theta_s$ and $\theta_x$, scalar-stepsize GGAMP can always be made to converge.}

\iftoggle{conference}{}{
Condition \eqref{eq:dampBnd} also helps to understand the effect of
$\kappa(\Abf)$ on the GGAMP convergence rate.
For example, if we equate $\theta_s=\theta_x =\theta$ for simplicity,
then \eqref{eq:dampBnd} implies that
\beq \label{eq:dampBnd2}
    \theta < \sqrt{C/\kappa(\Abf)}.
\eeq
Thus, if GGAMP converges at rate $\theta$, then
after $\theta$ is adjusted to ensure convergence,
GGAMP will converge at a rate below $\sqrt{C/\kappa(\Abf)}$.
So larger peak-to-average ratios $\kappa(\Abf)$ will result in slower convergence.
}

\subsection{Examples of Matrices}
\iftoggle{conference}{
To illustrate how the level of damping is affected by the nature of the matrix
$\Abf$, the full paper \cite{RanSchFle:14} evaluates the convergence
condition in several examples:
\begin{itemize}
\item \emph{i.i.d.\ matrices}:  These are shown to converge with no damping
(i.e.\ $\theta_s=\theta_x=1$) due to the Marcenko-Pastur Theorem.
\item \emph{Subsampled unitary matrices:}  These matrices have
a peak-to-average ratio $\kappa(\Abf)=1$ and hence also converge with no damping.
\item \emph{Linear filtering:}
When $\Abf$ implements
a circular convolution with filter $\hbf$, the ratio $\kappa(\Abf)$ can be
computed in frequency-domain as
$\kappa(\Abf) = \|\Fbf\hbf\|^2_\infty/\|\Fbf\hbf\|^2_2$
with discrete Fourier transform  matrix $\Fbf$.
Hence, filters with narrow bandwidths will require significant damping.
\item \emph{Walk-summable matrices:}  It is shown in \cite{RanSchFle:14} that
the walk-summability condition \cite{malioutov2006walk} that is sufficient
for Gaussian BP to converge is also sufficient for Gaussian GAMP.
\end{itemize}
}
{

To illustrate how the level of damping is affected by the nature of the matrix
$\Abf$, we consider several examples.

\paragraph{Large i.i.d. matrices}
Suppose that $\Abf \in \R^{m\x n}$ has i.i.d.\ components with zero
mean and unit variance.
For these matrices, we know from the rigorous state evolution analysis
\cite{BayatiM:11,javanmard2013state,BayLelMon:15} that, in the large-system limit
(i.e., $m,n\rightarrow\infty$ with fixed $m/n$), scalar-stepsize GGAMP will
converge without any damping.
We can reproduce this result using our analysis as follows:
By the Marcenko-Pastur Theorem~\cite{MarcenkoP:67}, it can be easily shown that
\begin{align}
    \kappa(\Abf)
    &\approx \frac{\min\{m,n\}}{m}\left[ 1 + \sqrt{\frac{m}{n}} \right]^2 \nonumber \\
    &\leq \frac{2\min\{m,n\}(m+n)}{mn}
\label{eq:kappaIid} ,
\end{align}
with equality when $m=n$, and
where the approximation becomes exact in the large-system limit.
Because this Marcenko-Pastur bound coincides with the $\theta_s = 1 = \theta_x$
case \eqref{eq:kappaNoDamp} of the convergence condition \eqref{eq:kappaCond},
our analysis implies that, for large i.i.d.\ matrices,
scalar stepsize GGAMP will converge without damping,
thereby confirming the state evolution analysis.
Note that we require that the asymptotic value of $m/n \neq 1$ so that
the inequality in \eqref{eq:kappaIid} is strict;
when $m=n$, \eqref{eq:kappaIid} becomes an equality and
we obtain a condition $\Gamma(\theta_s,\theta_x) = \|\Abf\|_2^2/\|\Abf\|_F^2$
right on the boundary between convergence and divergence,
where Theorem~\ref{thm:linConvUnif} does not make any
statements.

\paragraph{Subsampled unitary matrices}
Suppose that $\Abf$ is constructed by removing
either columns or rows, but not both, from a unitary matrix.
Then, $\kappa(\Abf)=1$, so, from \eqref{eq:kappaBnd},
$\kappa(\Abf) < \kappamax$ for any $\theta_s,\theta_x\in(0,1]$.
Hence, scalar GGAMP will converge with or without damping.

\paragraph{Linear filtering}
Suppose that $\Abf \in \R^{n \x n}$ is circulant with first column $\hbf$,
so that $(\Abf\xbf)_i = (\hbf \ast \xbf)_i$,
where $\ast$ denotes circular convolution.
(Linear convolution could be implemented via zero padding.)
Then, it can be shown that
\beq \label{eq:kappaH}
   \kappa(\Abf) =
   	\frac{\max_{k=0,\ldots,n-1} |H(e^{j2\pi k/n})|^2}
   	{\frac{1}{n}\sum_{k=0}^{n-1} |H(e^{j2\pi k/n})|^2} ,
\eeq
where $H(e^{j\omega})$ is the DTFT of $\hbf$.
Equation \eqref{eq:kappaH} implies that more damping is needed as
the filter becomes more narrowband.
For example, if $H(e^{j\omega})$ has a normalized
bandwidth of $B \in (0,1]$, then $\kappa(\Abf) \approx 1/B$ and,
relative to an allpass filter, GGAMP will need to slow by a factor of $O(\sqrt{B})$.

\paragraph{Low-rank matrices}
Suppose that $\Abf\in\R^{m\times n}$ has only
$r$ non-zero singular values, all of equal size.
Then
\[
    \kappa(\Abf) = \frac{\min\{m,n\}}{r},
\]
which, from \eqref{eq:dampBnd2}, implies the need to choose a damping constant
$\theta < \sqrt{C r/\min\{m,n\}}$, slowing the algorithm
by a factor of $\sqrt{\min\{m,n\}/r}$ relative to a full-rank matrix.
Hence, more damping is needed as the relative rank decreases.

\paragraph{Walk-summable matrices}  Closely related to Gaussian GAMP
is Gaussian belief propagation \cite{bickson2008gaussian,malioutov2006walk,weiss2000correctness}, which performs a similar
iterative algorithm to minimize a general quadratic
function of the form $f(\xbf) = \xbf\herm\Jbf\xbf + \mbox{Real}\{\cbf\herm\xbf\}$
for some positive definite matrix $\Jbf$.
\textb{Sufficient conditions for the convergence of Gaussian belief propagation were first shown in~\cite{weiss2000correctness,rusmevichientong2001analysis}, but those conditions are difficult to verify.
In a now classic result, \cite{malioutov2006walk} showed}
that Gaussian belief propagation will converge when
\beq \label{eq:walkSum}
    \lambda_{\max}\left( |\Ibf-\Jbf| \right) < 1, \mbox{ and } J_{ii}=1 \mbox{ for all } i,
\eeq
where $|\Ibf-\Jbf|$ is the component-wise magnitude.  The condition \eqref{eq:walkSum}
is called \emph{walk summability}, with the constraints $J_{ii}=1$ being for normalization.\par

\textb{ A quadratic function $f$ is said to be \emph{convex decomposable} if it can be written in the form $f(\xbf)=\sum_if_i(x_i) + \sum_{i,j}f_{ij}(x_i,x_j)$ where $\{f_i\}$ are strictly convex quadratic functions and $\{f_{ij}\}$ are convex quadratic functions. Moallemi and Van Roy~\cite{moallemi2009convergence} showed that if a quadratic objective function is convex decomposable then min-sum message passing converges to the global minimum. In~\cite{MalioutovJW:06}, it was shown that a function is convex decomposable if and only if it is walk-summable (i.e., the two properties are equivalent).
}\par
To compare walk summability with GGAMP, first observe that,
in the identical-variance case \eqref{eq:constVar}, GGAMP performs the same quadratic
minimization with a particular $\cbf$ and with
\[
    \Jbf = \tau_0\Abf\herm\Abf +\nu_w^{-1}\Ibf.
\]
Now, consider the high-SNR case, where $\tau_0=1$ and $\nu_w^{-1} \approx 0$, so that
$\Jbf \approx \Abf\herm\Abf$.
Then the walk-summability condition \eqref{eq:walkSum} reduces to
\beq \label{eq:walkSum2}
    \lambda_{\max}\big(|\Ibf-\Abf\herm\Abf| \big) < 1,
\eeq
where the normalizations $J_{ii}=1$ imply that the
columns of $\Abf$ have unit norm, i.e., that $\|\Abf\|^2_F=n$.
Note that, if \eqref{eq:walkSum2} is satisfied, then
\beqan
    \|\Abf\|^2_2 &=& \lambda_{\max}(\Abf\herm\Abf) \leq 1 + |1- \lambda_{\max}(\Abf\herm\Abf)| \\
    &=& 1 + |\lambda_{\max}(\Abf\herm\Abf - \Ibf)| \leq 1 + \lambda_{\max}\big(|\Abf\herm\Abf - \Ibf| \big) \\
    &=& 1 + \lambda_{\max}\big(|\Ibf - \Abf\herm\Abf| \big) < 2.
\eeqan
Applying these results to the $\kappa(\Abf)$ definition \eqref{eq:kappa}, we find
\beq
    \kappa(\Abf)
    = \frac{\|\Abf\|^2_2}{\|\Abf\|^2_F/\min\{m,n\}}
    < \frac{2\min\{m,n\}}{n}
    < \kappamaxnodamp ,
\eeq
where the latter inequality follows from inspection of \eqref{eq:kappaNoDamp}.
We conclude that, in the high-SNR regime,
walk summability is sufficient for GGAMP to converge with or without damping.\par
}

\section{Local Stability for Strictly Convex Functions} \label{sec:localStab}
We next consider the convergence with a
more general class of scalar estimation functions
$\gsbar$ and $g_x$: those that are twice continuously differentiable
with first derivatives bounded as
\beq \label{eq:gderivbnd}
    [\gsbar'(\pbfbar,\nubf_p)]_i \in (0,1), \quad
    [g_x'(\rbf,\taubf_r)]_j \in (0,1),
\eeq
for all $\pbfbar$, $\rbf$, $\nubf_p$ and $\taubf_r$.
This condition arises in the important case of minimizing
strictly convex functions.
Specifically, if GAMP is used in max-sum mode so that the scalar
estimation functions are given by  \eqref{eq:gxMAP} and \eqref{eq:gsbarMAP}
with strictly convex, twice differentiable functions $G_i$
and $F_j$, then \eqref{eq:gxMAP}, \eqref{eq:gsbarMAP},
and \eqref{eq:prox'} show that
the conditions in \eqref{eq:gderivbnd} will be satisfied.

\bluecolor
\medskip
\begin{definition}
        Let $\xbf^{t+1}=\bm{f}_t(\xbf^t)$ for $t=0,1,2,\cdots$ be a dynamical system with a fixed point $\xbf^{*}$ (i.e., $\bm{f}_t(\xbf^{*})=\xbf^{*}~\forall t$). We say that the system is \emph{locally stable} at $\xbf^{*}$ if $\exists \delta>0$ such that, if $\|\xbf^{0}-\xbf^{*}\|<\delta$, then $\lim_{t\rightarrow\infty}\xbf^{t}=\xbf^{*}$.
\end{definition}
\medskip
\color{black}

Outside of the Gaussian scenario, we have not yet established conditions on
the global convergence of GAMP for general scalar estimation functions.\footnote{\textb{Interestingly, it was shown by Moallemi and Van Roy~\cite{moallemi2010convergence} that, for a certain class of convex optimization problems characterized by ``scaled diagonal dominance'', max-sum BP converges. As future work, it would be interesting to study whether max-sum GAMP also converges for this class of problems.}}
Instead, we now establish conditions on \emph{local} stability,
as defined in \cite{Vidyasagar:78}.
To simplify the analysis, we will assume that the GAMP algorithm
uses arbitrary but \emph{fixed} stepsize vectors $\nubf_p$ and $\taubf_r$.

Under these assumptions, consider any fixed point $(\pbfbar,\rbf)$ of the GAMP method,
and define the matrices
\begin{subequations} \label{eq:QsxDeriv}
\beqa
    \Qbf_s &:=& \diag(\qbf_s), \quad \qbf_s := \gsbar'(\pbfbar,\nubf_p), \\
    \Qbf_x &:=& \diag(\qbf_x), \quad \qbf_x := g'_x(\rbf,\taubf_r),
\eeqa
\end{subequations}
evaluated at that fixed point.  Note that, under assumption \eqref{eq:gderivbnd},
the components of $\qbf_s$ and $\qbf_x$ lie in $(0,1)$.
Define the matrix
\beq \label{eq:Anorm}
    \tilde{\Abf} := \diag{}^{1/2}(\nubf_p.\qbf_s)\Abf\diag{}^{1/2}(\taubf_r.\qbf_x).
\eeq
\iftoggle{conference}{
For reasons explained in the full paper \cite{RanSchFle:14}, we will refer to this
matrix as the \emph{row-column normalized matrix}.

Observe that if $\nubf_p$ and $\taubf_r$ were
fixed points of the GAMP-adapted stepsizes,
then lines~\ref{line:taupt}, \ref{line:taust}, \ref{line:taurt}, and \ref{line:tauxt}
in Algorithm~\ref{algo:gamp} would imply that
\beq \label{eq:taurpFix}
    \frac{1}{\nubf_p} = \Sbf\Qbf_x\taubf_r, \quad
    \frac{1}{\taubf_r} = \Sbf\tran\Qbf_s\nubf_p.
\eeq
The result below, however, applies
to \emph{arbitrary} fixed stepsize vectors $\nubf_p$ and $\taubf_r$---not
necessarily the ones satisfying \eqref{eq:taurpFix}.}{
Then \eqref{eq:QsxDeriv}-\eqref{eq:Anorm}, together with lines~\ref{line:taust}
and \ref{line:taurt} of Algorithm~\ref{algo:gamp}, imply
\beqa
    \sum_{i=1}^m |\tilde{A}_{ij}|^2 &=&
    q_{x_j}\tau_{r_j} \sum_{i=1}^m \nu_{p_i}q_{s_i}|A_{ij}|^2  \\
    &=& q_{x_j}\tau_{r_j} \sum_{i=1}^m S_{ij}\nu_{p_i}q_{s_i} = q_{x_j} < 1.
    \label{eq:AnormCol}
\eeqa
Hence, the column norms of $\tilde{\Abf}$ in \eqref{eq:Anorm}
are less than one.
Similar arguments can be use to establish that, for any $i$,
\beqa
    \sum_{j=1}^n |\tilde{A}_{ij}|^2 =  q_{s_i} < 1,
    \label{eq:AnormRow}
\eeqa
so that $\tilde{\Abf}$ also has row norms less than one.
We will thus call $\tilde{\Abf}$ the \emph{row-column normalized matrix}.
}

\medskip
\begin{theorem} \label{thm:localStab}
Consider any fixed point $(\sbfbar,\xbf)$
of GAMP Algorithm~\ref{algo:gamp} or Algorithm~\ref{algo:gampUnif}
with \emph{fixed} vector or scalar stepsizes $\nubf_p$ and $\taubf_r$,
respectively, and scalar estimation functions $\gsbar$ and $g_x$ satisfying
the above conditions.  Then,
the fixed point is locally stable if
\beq \label{eq:sigBnd}
    \theta_s\theta_x\|\tilde{\Abf}\|_2^2 < 1,
\eeq
for $\tilde{\Abf}$ defined in \eqref{eq:Anorm}.
For the Gaussian GAMP algorithm, the same condition implies the algorithm
is globally stable.
\end{theorem}
\iftoggle{conference}{}{
\begin{IEEEproof} See Appendix~\ref{sec:localStabPf}.
\end{IEEEproof}
}

\medskip
To relate this condition to Theorem~\ref{thm:linConvUnif},
consider the case when $\nubf_s$ and $\taubf_x$ are fixed points
of \eqref{eq:tausxFix} with  $\Sbf=\Abf.\Abf$, i.e., the component-wise magnitude square of $\Abf$.
\iftoggle{conference}{In this case, the full paper \cite{RanSchFle:14} shows
that a sufficient condition to satisfy \eqref{eq:sigBnd} is
\beq \label{eq:kappaCondVec}
    \kappa(\tilde{\Abf}) \leq \frac{1}{\theta_x\theta_s\max\{\qbar_s,\qbar_x\}},
\eeq
where $\qbar_s$ and $\qbar_x \in (0,1)$ are the average values of the components
of $\qbf_s$ and $\qbf_x$ respectively.}{
From \eqref{eq:AnormCol} and \eqref{eq:AnormRow}, we have that
\[
    \|\tilde{\Abf}\|^2_F = m\qbar_s = n\qbar_x \leq \min\{m,n\}\max\{\qbar_s,\qbar_x\},
\]
where
\[
    \qbar_s = \frac{1}{m}\sum_{i=1}^m q_{s_i}, \quad
    \qbar_x = \frac{1}{n}\sum_{j=1}^n q_{x_j}.
\]
Thus, the peak-to-average ratio of $\tilde{\Abf}$
as defined in \eqref{eq:kappa} is bounded below as
\[
    \kappa(\tilde{\Abf}) \geq \frac{\|\tilde{\Abf}\|^2_2}{\max\{\qbar_s,\qbar_x\}}.
\]
Hence, a sufficient condition to satisfy \eqref{eq:sigBnd} is given by
\beq \label{eq:kappaCondVec}
    \kappa(\tilde{\Abf}) < \frac{1}{\theta_x\theta_s\max\{\qbar_s,\qbar_x\}}.
\eeq
}
In comparison, \eqref{eq:kappaCond} and \eqref{eq:kappaBnd} show that
a Gaussian GAMP with scalar step sizes converges is
$\kappa(\Abf) < C/(\theta_s\theta_x)$.
We conclude that the sufficient condition for the vector-stepsize GAMP
algorithm to converge is similar to the scalar-stepsize GAMP algorithm,
but where the peak-to-average ratio is measured on a certain normalized matrix.

\section{Numerical Results}

In this section, we present some numerical simulations to verify Theorems~\ref{thm:linConvUnif} and~\ref{thm:localStab}.
This section is divided into two parts: the first part is on the global convergence of damped GGAMP (Theorem~\ref{thm:linConvUnif}) and the second part is on the local stability of damped GAMP (Theorem~\ref{thm:localStab}).

For both experiments, we first generated a matrix $\Rbf\in\mathbb{R}^{m\times n}$ with elements drawn i.i.d.\ ${\mathcal N}(0,1)$ and computed its SVD to get orthogonal matrices $\Ubf,\Vbf$ such that $\Rbf=\Ubf\bm{\Lambda}\Vbf\tran$.
Then we set $\Abf=\Ubf\bm{\Sigma}\Vbf\tran$ for $\bm{\Sigma}=\diag\{\sigma_1,...,\sigma_{r}\}$, where $r=\min\{m,n\}$, $\sigma_1=1$, and $\sigma_i/\sigma_{i-1}=\rho~\forall i$.
The value of $\rho$ was chosen to achieve a desired value of the peak-to-average ratio of the squared singular values of $\Abf$, i.e., $\kappa(\Abf)$ in \eqref{eq:kappa}.
Finally, the measurements $\ybf$ were generated according to $\ybf = \Abf\xbf + \wbf$ for the AWGN case, or $\ybf = \text{sign}(\Abf\xbf + \wbf)$ for the binary case, where in either case $\wbf$ was a realization of white Gaussian noise.
The variance of $\wbf$ was chosen to achieve an SNR of $50$ dB, where SNR $:=\Exp\{\|\Abf\xbf\|^2\}/\Exp\{\|\wbf\|^2\}$.

\subsection{Global Convergence of Damped GGAMP}

In this experiment, the elements of $\xbf$ were drawn i.i.d. ${\mathcal N}(0,1)$ and the measurements were generated using the AWGN model as discussed above.
For each choice of damping factor $\theta_s=\theta_x$, scalar stepsize GGAMP was run from the fixed initialization $\{\xbf^0\!=\!\bm{0}$, $\sbfbar^{-1}\!=\!\bm{0}$, $\tau_x\!=\!1\}$ and the MSE after \textb{$5000$} iterations was recorded.
This experiment was then repeated for \textb{$100$} realizations of $\{\Abf,\xbf,\wbf\}$.
The damping factors $\theta_s=\theta_x$ were varied from $0.7$ to $1$ in steps of $0.005$.
To test the validity of Theorem~\ref{thm:linConvUnif}, we present the results in term of the ``excess MSE,'' defined as the ratio of the MSE achieved by GAMP to the MMSE, which was computed in closed form.
\textb{To enhance the readability of the plots, the excess MSE was clipped at $100$~dB.}

\begin{figure}[t]
\centering
\psfrag{y}[Bc][c][\sz]{\sf excess MSE [dB]}
\psfrag{x}[tc][c][\sz]{$\kappamax$}
\psfrag{kappaA}[c][c][0.65]{$\kappa(\Abf)$}
\psfrag{GAMP}[tc][tc][0.65]{GGAMP}
  \includegraphics[scale=0.45]{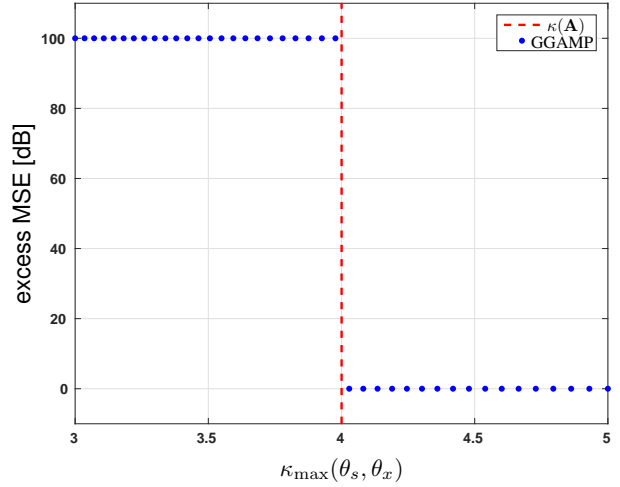}
  \caption{The excess MSE of GGAMP vs $\kappamax$ for $\kappa(\Abf)=4$. Each point represents one realization, \textb{and excess MSE values were clipped at $100$~dB.  To the right of the red dashed line, the condition $\kappamax>\kappa(\Abf)$ is satisfied, in which case GGAMP converges to the MMSE solution, as predicted by Theorem~\ref{thm:linConvUnif}.}}
  \label{fig:gauss_awgn_plot1}
\end{figure}

\begin{figure}[t]
\centering
\psfrag{y}[Bc][c][\sz]{\sf excess MSE [dB]}
\psfrag{x}[tc][c][\sz]{$\kappamax$}
\psfrag{kappaA}[c][c][0.65]{$\kappa(\Abf)$}
\psfrag{GAMP}[tc][tc][0.65]{GGAMP}
  \includegraphics[scale=0.45]{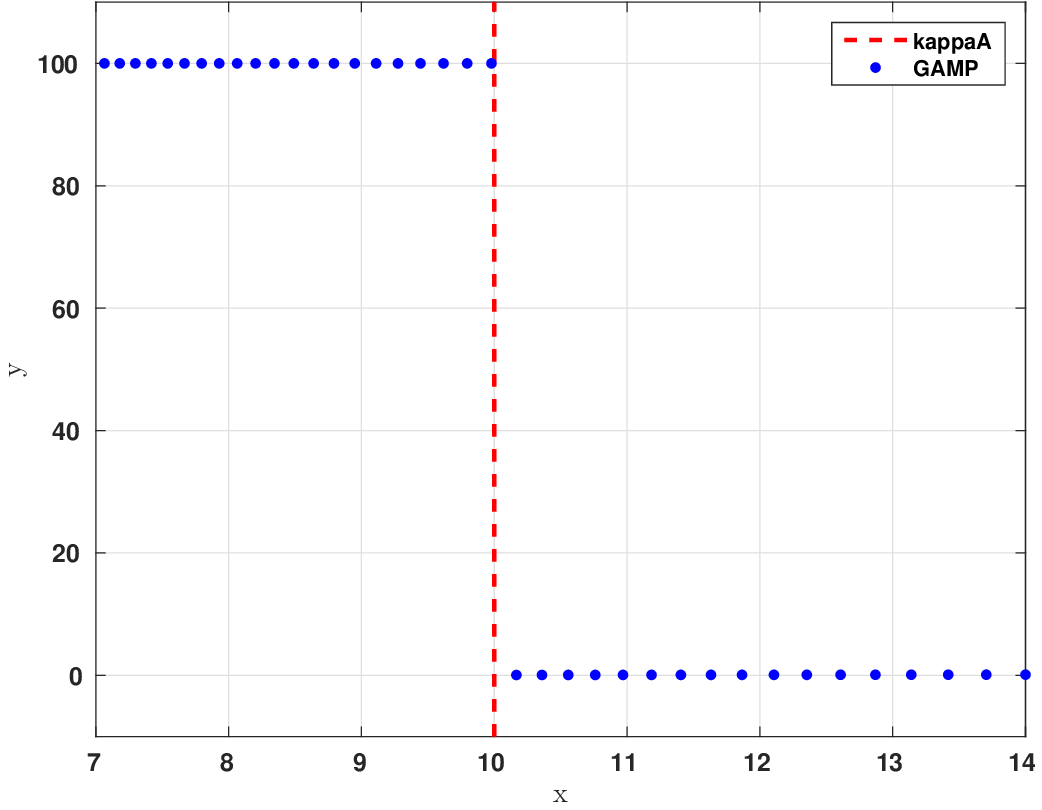}
  \caption{The excess MSE of GGAMP vs $\kappamax$ for $\kappa(\Abf)=10$. Each point represents one realization, \textb{and excess MSE values were clipped at $100$~dB.  To the right of the red dashed line, the condition $\kappamax>\kappa(\Abf)$ is satisfied, in which case GGAMP converges to the MMSE solution, as predicted by Theorem~\ref{thm:linConvUnif}.}}
  \label{fig:gauss_awgn_plot2}
\end{figure}

Figures~\ref{fig:gauss_awgn_plot1} and~\ref{fig:gauss_awgn_plot2} show the excess MSE versus $\kappamax$, which---according to Theorem~\ref{thm:linConvUnif}---is the maximum allowed value of $\kappa(\Abf)$ under which GGAMP will converge with damping factors $(\theta_s,\theta_x)$, as defined in \eqref{eq:kappaCond}.
In both figures, the dimensions of $\Abf$ were $200\times100$, and the excess MSE from each realization is plotted as a dot.
The figures show that the excess MSE was zero~dB whenever $\kappamax>\kappa(\Abf)$,
and conversely the excess MSE was greater than zero~dB whenever $\kappamax<\kappa(\Abf)$,
which verifies the claim of Theorem~\ref{thm:linConvUnif}.

\subsection{Local Convergence of GAMP}

\begin{figure}[t]
\centering
\psfrag{y}[Bc][c][\sz]{\sf excess MSE [dB]}
\psfrag{x}[tc][c][\sz]{$\theta_s\theta_x\|\tilde{\Abf}\|_2^2$}
\psfrag{kappaA}[c][c][0.65]{$\kappa(\Abf)$}
\psfrag{GAMP}[tc][tc][0.65]{GAMP}
  \includegraphics[scale=0.45]{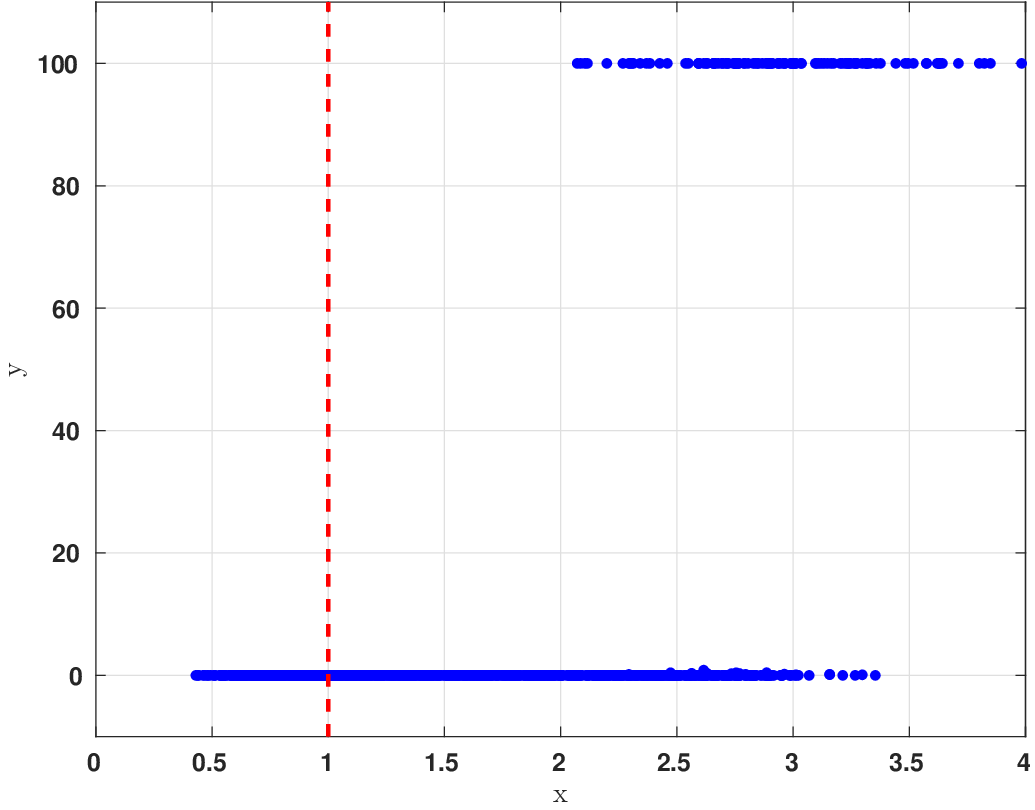}
  \caption{Excess MSE (dB) vs $\theta_s\theta_x\|\tilde{\Abf}\|_2^2$ for BG prior and AWGN likelihood and $\kappa(\Abf)=4$. \textb{Excess MSE values were clipped at $100$~dB.  To the left of the red dashed line, the sufficient condition $\theta_s\theta_x\|\tilde{\Abf}\|_2^2<1$ is satisfied, in which case damped GAMP locally converges to a fixed point, as predicted by Theorem~\ref{thm:localStab}}.}
  \label{fig:bg_awgn_plot1}
\end{figure}

\begin{figure}[t]
\centering
\psfrag{y}[Bc][c][\sz]{\sf excess MSE [dB]}
\psfrag{x}[tc][c][\sz]{$\theta_s\theta_x\|\tilde{\Abf}\|_2^2$}
\psfrag{kappaA}[c][c][0.65]{$\kappa(\Abf)$}
\psfrag{GAMP}[tc][tc][0.65]{GAMP}
 \includegraphics[scale=0.45]{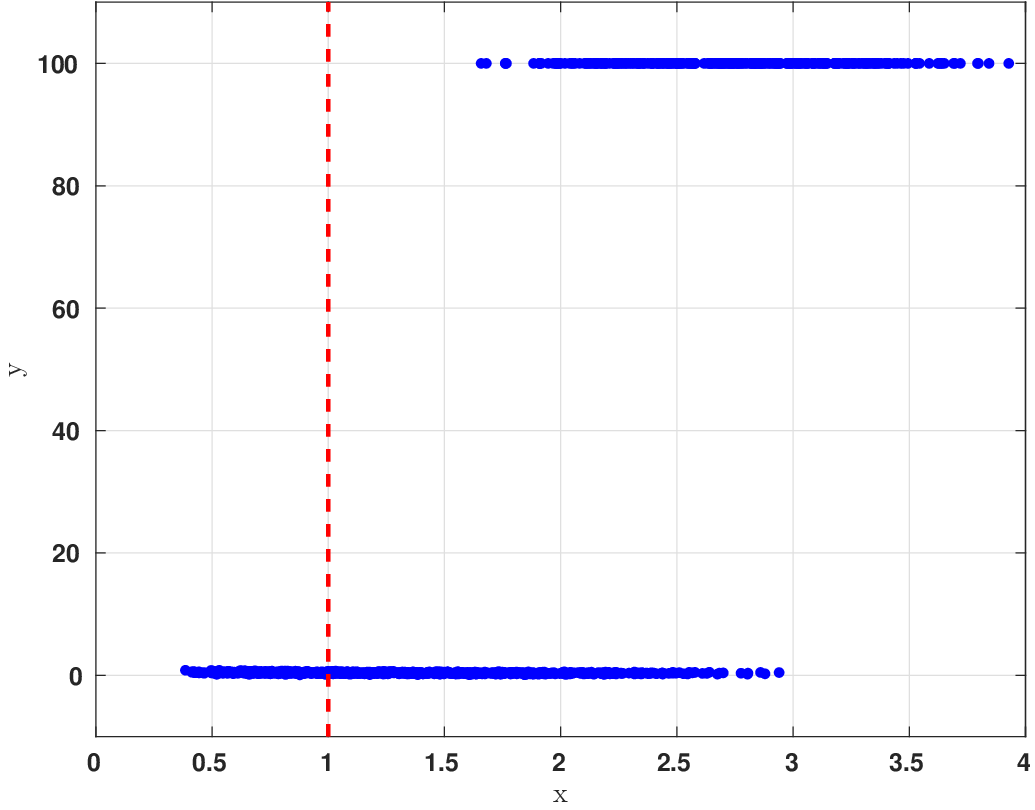}
  \caption{Excess MSE (dB) vs $\theta_s\theta_x\|\tilde{\Abf}\|_2^2$ for BG prior and AWGN likelihood and $\kappa(\Abf)=10$. \textb{Excess MSE values were clipped at $100$~dB.  To the left of the red dashed line, the sufficient condition $\theta_s\theta_x\|\tilde{\Abf}\|_2^2<1$ is satisfied, in which case damped GAMP locally converges to a fixed point, as predicted by Theorem~\ref{thm:localStab}.}}
  \label{fig:bg_awgn_plot2}
\end{figure}

\begin{figure}[t]
\centering
\psfrag{y}[Bc][c][\sz]{\sf excess MSE [dB]}
\psfrag{x}[tc][c][\sz]{$\theta_s\theta_x\|\tilde{\Abf}\|_2^2$}
\psfrag{kappaA}[c][c][0.65]{$\kappa(\Abf)$}
\psfrag{GAMP}[tc][tc][0.65]{GAMP}
  \includegraphics[scale=0.45]{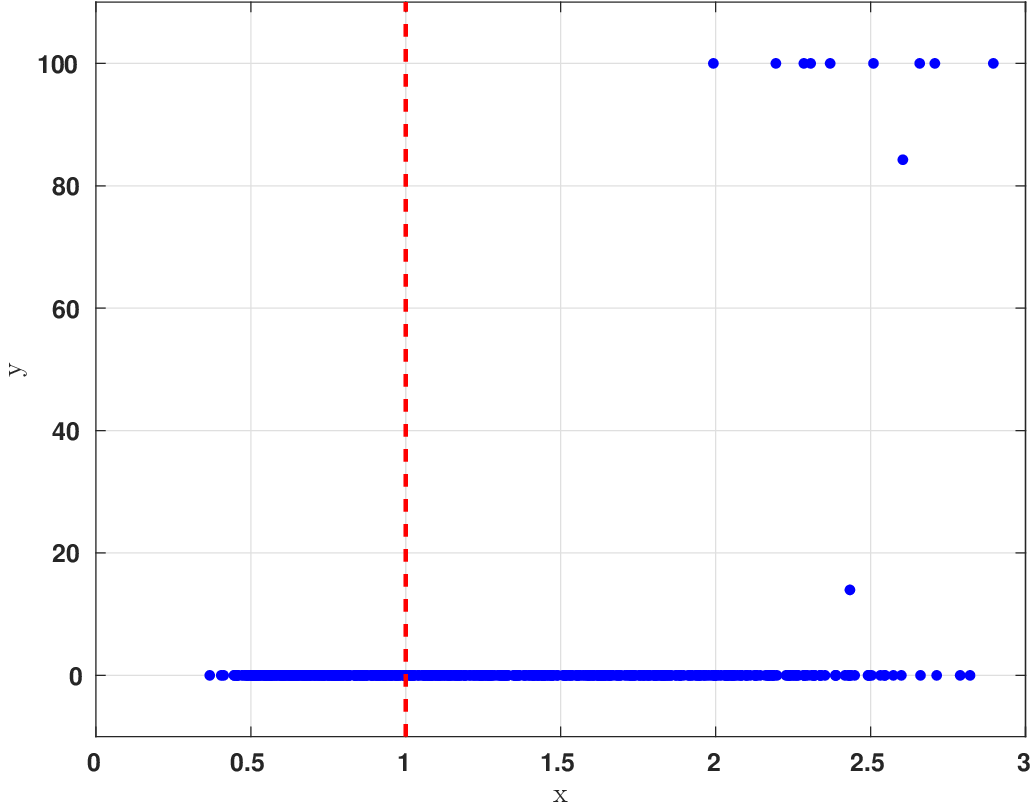}
  \caption{Excess MSE (dB) vs $\theta_s\theta_x\|\tilde{\Abf}\|_2^2$ for BG prior and Probit likelihood and $\kappa(\Abf)=4$. \textb{Excess MSE values were clipped at $100$~dB.  To the left of the red dashed line, the sufficient condition $\theta_s\theta_x\|\tilde{\Abf}\|_2^2<1$ is satisfied, in which case damped GAMP locally converges to a fixed point, as predicted by Theorem~\ref{thm:localStab}.}}
  \label{fig:bg_binary_plot1}
\end{figure}

\begin{figure}[t]
\centering
\psfrag{y}[Bc][c][\sz]{\sf excess MSE [dB]}
\psfrag{x}[tc][c][\sz]{$\theta_s\theta_x\|\tilde{\Abf}\|_2^2$}
\psfrag{kappaA}[c][c][0.65]{$\kappa(\Abf)$}
\psfrag{GAMP}[tc][tc][0.65]{GAMP}
  \includegraphics[scale=0.45]{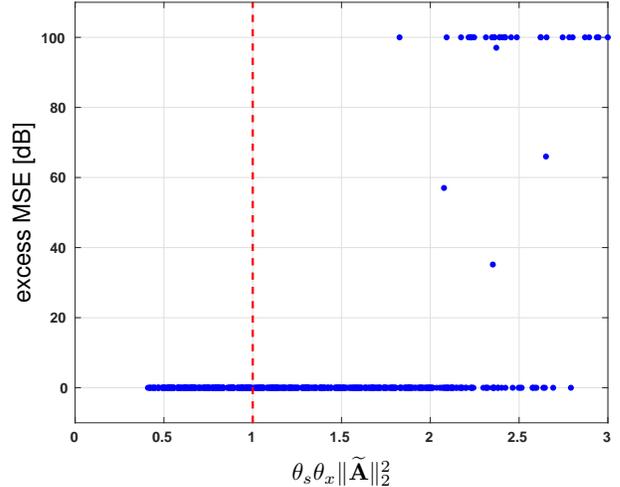}
  \caption{Excess MSE (dB) vs $\theta_s\theta_x\|\tilde{\Abf}\|_2^2$ for BG prior and Probit likelihood and $\kappa(\Abf)=10$. \textb{Excess MSE values were clipped at $100$~dB.  To the left of the red dashed line, the sufficient condition $\theta_s\theta_x\|\tilde{\Abf}\|_2^2<1$ is satisfied, in which case damped GAMP locally converges to a fixed point, as predicted by Theorem~\ref{thm:localStab}.}}
  \label{fig:bg_binary_plot2}
\end{figure}

To test the local stability of damped GAMP, we used the following procedure.
For each realization of $\{\Abf,\xbf,\ybf\}$, the parameters $\{\nubf_p,\taubf_r,\theta_s,\theta_x\}$ were chosen and vector-stepsize GAMP was run from the initialization $\{\xbf^0=\bm{0}$, $\sbfbar^{-1}=\bm{0}$, $\taubf_x=\bm{1}\}$ with the stepsizes fixed at the chosen $\{\nubf_p,\taubf_r\}$.
The values of $\{\nubf_p,\taubf_r,\theta_s,\theta_x\}$ were chosen so that GAMP converged to some fixed point $\{\xbf, \sbfbar,\pbfbar,\rbf\}$; more details are provided below.
Next, GAMP was initialized near to the fixed point and tested for local convergence (under the same fixed stepsizes $\{\nubf_p,\taubf_r\}$.)
In particular, it was initialized at $\{\xbf^0=\xbf+\xbf_{\varepsilon}$, $\sbfbar^{-1}=\sbfbar\}$, where the elements of $\xbf_{\varepsilon}$ were drawn i.i.d.\ ${\mathcal N}(0,1)$, with $\xbf_{\varepsilon}$ subsequently normalized such that the initial MSE was $15$~dB above the MSE at the fixed point.
This test was repeated $20$ times for each fixed point.
If $\theta_s\theta_x\|\tilde{\Abf}\|_2^2<1$ then, according to Theorem~\ref{thm:localStab}, GAMP should converge to the fixed point.
Each dot in Figures~\ref{fig:bg_awgn_plot1}-\ref{fig:bg_binary_plot2} represents the excess MSE, now defined as the ratio of the \emph{maximum} MSE among all local runs of GAMP to the MSE at the fixed point.
The above procedure was repeated for a range of $\theta_s=\theta_x$ and many realizations of $\{\Abf,\xbf,\ybf\}$, as detailed below.
\textb{As before, the excess MSE values were clipped at $100$~dB before plotting.}

Figures~\ref{fig:bg_awgn_plot1} and~\ref{fig:bg_awgn_plot2} show the excess MSE versus $\theta_s\theta_x\|\tilde{\Abf}\|_2^2$ for Bernoulli-Gaussian $\xbf$ \textb{with sparsity rate 0.1} and AWGN measurements.
Figure~\ref{fig:bg_awgn_plot1} investigates the case where $\kappa(\Abf)=4$ and Figure~\ref{fig:bg_awgn_plot2} investigates the case where $\kappa(\Abf)=10$.
For each plot,
the dimensions of $\Abf$ were $200\times100$,
the stepsizes were $\nu_{p_i}=(\sum_{j=1}^n A_{ij}^2)^{-1}~\forall i$,
the damping factors $\theta_s=\theta_x$ were varied from $0.45$ to $0.95$ in steps of $0.05$,
and \textb{$50$} realizations of $\{\Abf,\xbf,\ybf\}$ were tested.
Also, $\tau_{r_j}=(\sum_{i=1}^m A_{ij}^2)^{-1}$ in Figure~\ref{fig:bg_awgn_plot1} and $\tau_{r_j}=(10\sum_{i=1}^m A_{ij}^2)^{-1}$ in Figure~\ref{fig:bg_awgn_plot2}, for all $j$.
This particular choice of $\taubf_r$ was used to ensure that the fixed-stepsized GAMP converged to a fixed point for the chosen range of $\theta_s,\theta_x$.

Figure~\ref{fig:bg_binary_plot1} and~\ref{fig:bg_binary_plot2} show the excess MSE versus $\theta_s\theta_x\|\tilde{\Abf}\|_2^2$ for Bernoulli-Gaussian $\xbf$ \textb{with sparsity rate 0.1} and binary measurements.
Figure~\ref{fig:bg_binary_plot1} investigates the case where $\kappa(\Abf)=4$ and Figure~\ref{fig:bg_binary_plot2} investigates the case where $\kappa(\Abf)=10$.
For each plot,
the dimensions of $\Abf$ were $400\times100$,
the stepsizes were $\nu_{p_i}=10~\forall i$ and $\tau_{r_j}=1~\forall j$,
the damping factors $\theta_s=\theta_x$ were varied from $0.45$ to $0.95$ in steps of $0.05$,
and $50$ realizations of $\{\Abf,\xbf,\ybf\}$ were tested.

Figures~\ref{fig:bg_awgn_plot1}-\ref{fig:bg_binary_plot2} show an excess MSE of $\approx 0$~dB whenever $\theta_s\theta_x\|\tilde{\Abf}\|_2^2<1$, hence verifying Theorem~\ref{thm:localStab}.

\section*{Conclusions}
A key outstanding issue for the adoption of
AMP-related methods is their convergence for generic finite-dimensional
linear transforms.
Similar to other loopy BP-based methods, standard forms of AMP
may diverge.
In this paper, we presented a damped version of the generalized AMP algorithm
that, when used with fixed stepsizes, can guarantee global convergence for
Gaussian distributions
and local convergence for the minimization of strictly convex functions
(i.e., strictly concave log-priors).
The required amount of damping is related to the peak-to-average ratio
of the squared singular values of the transform matrix.
However, much remains unanswered:
Most importantly, we have yet to derive a condition for global
convergence even in the case of strictly convex functions.
Secondly, our analysis assumes the use of fixed stepsizes.
Third, short of
computing the peak-to-average singular-value ratio,
we proposed no method to compute the damping constants.
Hence, an adaptive method may be useful in practice.
One such method, \cite{Vila:ICASSP:15}, has been proposed, but
it comes without convergence guarantees.
Thus, future work might aim to analyze the convergence of such methods.
\textb{
Also, a more recent algorithm, Vector AMP (VAMP) \cite{rangan2017vamp,fletcher2017learning},
has improved convergence on larger classes of random matrices.
Another line of future work could seek conditions for convergence of VAMP
on deterministic matrices.
}

\iftoggle{conference}{}{
\appendices
\section{Proof of Theorem \ref{thm:varConv}}
\label{sec:varConvPf}

The variance updates of both Algorithms~\ref{algo:gamp} and
\ref{algo:gampUnif} are both of the form \eqref{eq:tausxFix} with
different choices of $\Sbf$. So, the theorem will be proven by
showing  that the updates \eqref{eq:tausxFix} converge for any
non-negative matrix $\Sbf \geq 0$.  To this end,
we use the results in \cite{yates:95}.
Specifically, for any $\nubf_w$ and $\taubf_0>0$,
define the functions
\beqan
    \Phi_s(\taubf_x) &:=& \left[ \Sbf\taubf_x +
     1./\nubf_w\right]^{-1} \\
    \Phi_x(\nubf_s) &:=& \left[
        \Sbf\tran\nubf_s + 1./\taubf_0 \right]^{-1}
\eeqan
so that the updates \eqref{eq:tausxFix} can be written as
\[
    \nubf_s^t = \Phi_s(\taubf_x^t), \quad
    \taubf_x^{\tp1} = \Phi_x(\nubf^t_s).
\]
It is easy to check that, for any $\Sbf \geq 0$,
\begin{enumerate}
\item[(i)] $\Phi_s(\taubf_x) > 0$,
\item[(ii)] $\taubf_x \geq \taubf_x' \Rightarrow
    \Phi_s(\taubf_x) \leq     \Phi_s(\taubf_x')$, and
\item[(iii)] For all $\alpha > 1$,
    $\Phi_s(\alpha\taubf_x) > (1/\alpha)\Phi_s(\taubf_x)$.
\end{enumerate}
with the analogous properties being satisfied by $\Phi_x(\nubf_s)$.
Now let $\Phi := \Phi_x\circ \Phi_s$ be the composition
of the two functions so that $\taubf^{\tp1}_x = \Phi(\taubf^t_x)$.
Then, $\Phi$ satisfies the three properties:
\begin{enumerate}
\item[(i)] $\Phi(\taubf_x) > 0$,
\item[(ii)] $\taubf_x \geq \taubf_x' \Rightarrow
    \Phi(\taubf_x) \geq  \Phi(\taubf_x')$, and
\item[(iii)] For all $\alpha > 1$,
    $\Phi(\alpha\taubf_x) < \alpha\Phi(\taubf_x)$.
\end{enumerate}
Also, for any $\nubf_s\geq 0$, we have $\Phi_x(\nubf_s) \leq \taubf_0$
and therefore, $\Phi(\taubf_x) \leq \taubf_0$ for all $\taubf_x\geq 0$.
Hence, taking any $\taubf_x \geq \taubf_0$, we obtain:
\[
    \taubf_x \geq \Phi(\taubf_x).
\]
\textb{Using Theorem 2 in \cite{yates:95}, it can be shown that the updates}
$\taubf^{\tp1}_x = \Phi(\taubf_x^t)$ converge to a unique fixed point.
A similar argument shows that $\nubf_s^t$ also converges
to a unique fixed point.

\section{Linear System Stability Condition}

The proofs of both Theorems~\ref{thm:linConvUnif} and \ref{thm:localStab}
are based on analyzing the GAMP algorithm via an equivalent
linear system and then applying results from linear stability theory.
For both results we will show that the condition of the theorem
is equivalent to an eigenvalue test on a certain matrix.

First consider the Gaussian GAMP algorithm with fixed vector stepsizes.
With fixed stepsizes and
Gaussian estimation functions \eqref{eq:gsxLin},
Algorithm~\ref{algo:gamp} reduces to a linear system:
\begin{subequations} \label{eq:gampLin}
\beqa
    \sbfbar^t &=& (1-\theta_s)\sbfbar^{\tm1} + \theta_s
        \Qbf_s( \sbfbar^{\tm1} + \nubf_p.\Abf\xbf^t)
        \nonumber \\ && \mbox{}
        - \theta_s\nubf_w.\ybf\\
    \xbf^{\tp1} &=& (1-\theta_x)\xbf^t + \theta_x
    \Qbf_x(\xbf^t - \taubf_r.\Abf\herm\sbfbar^t - \xbf_0)
        \nonumber \\ && \mbox{}
        + \theta_x\xbf_0,
\eeqa
\end{subequations}
where
\begin{subequations} \label{eq:Qsx}
\beqa
    \Qbf_s &=& \diag(\qbf_s), \quad \qbf_s = \nubf_w./(\nubf_w+\nubf_p), \\
    \Qbf_x &=& \diag(\qbf_x), \quad \qbf_x = \taubf_0./(\taubf_0+\taubf_r).
\eeqa
\end{subequations}
Note that the components of $\qbf_s$ and $\qbf_x$ are in $(0,1)$.
We can write the system \eqref{eq:gampLin} in matrix form as
\beq \label{eq:gampLinMat}
    \left[ \begin{array}{c} \sbfbar^t \\ \xbf^{\tp1} \end{array} \right]
    = \Gbf \left[ \begin{array}{c} \sbfbar^{\tm1} \\ \xbf^{t} \end{array} \right]  + \bbf,
\eeq
for an appropriate matrix $\Gbf$ and vector $\bbf$.  The matrix $\Gbf$ is given by
\beq \label{eq:Gdef}
    \Gbf :=
    \left[ \begin{array}{cc} I & 0 \\ -\theta_x\diag(\taubf_x)\Abf\herm &
        \Dbf_x  \end{array} \right]
    \left[ \begin{array}{cc} \Dbf_s & \theta_s\diag(\nubf_s)\Abf \\ 0 & I \end{array} \right],
\eeq
where
\begin{subequations} \label{eq:Dsxdef}
\beqa
     \Dbf_s &=& (1-\theta_s)\Ibf + \theta_s\Qbf_s \label{eq:Dxdef} \\
     \Dbf_x &=& (1-\theta_x)\Ibf + \theta_x\Qbf_x. \label{eq:Dsdef}
\eeqa
\end{subequations}
Here we have used that
\beq \label{eq:qtauprod}
    \qbf_s.\nubf_p = \nubf_s, \quad
    \qbf_x.\taubf_r = \taubf_x.
\eeq
Note that both $\Dbf_x$ and $\Dbf_s$ are diagonal matrices with entries
in the interval $(0,1)$.

Now, consider the case of the more general scalar estimation functions
satisfying \eqref{eq:gderivbnd}
and other assumptions in Section~\ref{sec:localStab}.
Due to the differentiability assumptions, to prove the local stability,
we only have to look at the
linearization of the system around the fixed points \cite{Vidyasagar:78}.
With fixed stepsizes,
the linearization of the updates in Algorithm~\ref{algo:gamp}
around any fixed point is given by
\begin{subequations} \label{eq:gampLinGen}
\beqa
    \sbfbar^t &=& (1-\theta_s)\sbfbar^{\tm1} + \theta_s
        \Qbf_s( \sbfbar^{\tm1} + \nubf_p.\Abf\xbf^t)\\
    \xbf^{\tp1} &=& (1-\theta_x)\xbf^t + \theta_x
    \Qbf_x(\xbf^t - \taubf_r.\Abf\herm\sbfbar^t)
\eeqa
\end{subequations}
where the matrices $\Qbf_s$ and $\Qbf_x$ in \eqref{eq:Qsx}
are replaced by the derivatives \eqref{eq:QsxDeriv}.
This linear system is also of the form \eqref{eq:gampLinMat} with the same
matrix \eqref{eq:Gdef}.
Also, under the assumptions of the theorem, $\qbf_s$ and $\qbf_x$
are vectors with components in $(0,1)$.

Hence, we conclude that to prove the global stability of
Gaussian GAMP,
or the local stability of GAMP under the assumptions
of Theorem~\ref{thm:localStab},
it suffices to show that the
linear system \eqref{eq:gampLinMat} with a
matrix $\Gbf$ of the form \eqref{eq:Gdef} is stable.
The matrices $\Dbf_s$ and $\Dbf_x$ are given in \eqref{eq:Dsxdef}
where $\Qbf_s$ and $\Qbf_x$ are diagonal matrices with elements
in $(0,1)$.

To evaluate this condition, first recall that
the linear system \eqref{eq:gampLinMat}
is stable when the eigenvalues of $\Gbf$ are in the unit circle.
However, if we define
\[
    \Tbf =  \left[ \begin{array}{cc} \diag^{-1/2}(\theta_s\nubf_s) & 0 \\
     0 & \diag^{-1/2}(\theta_x\taubf_x)  \end{array} \right],
\]
the eigenvalues of $\Gbf$ are identical to those of $\Hbf$ given by
\beq \label{eq:Hdef}
    \Hbf := \Tbf\Gbf\Tbf^{-1} =
    \left[ \begin{array}{cc} \Ibf & 0 \\ -\Fbf\herm & \Dbf_x  \end{array} \right]
    \left[ \begin{array}{cc} \Dbf_s & \Fbf \\ 0 & \Ibf \end{array} \right],
\eeq
where
\beq \label{eq:Fdef}
    \Fbf = \sqrt{\theta_s\theta_x}\diag(\nubf_s^{1/2})\Abf\diag(\taubf_x^{1/2}).
\eeq
Expanding the matrix product in \eqref{eq:Hdef}, we get
\beq \label{eq:Hprod}
    \Hbf = \left[ \begin{array}{cc} \Dbf_s & \Fbf \\
    -\Fbf\herm\Dbf_s & \Dbf_x-\Fbf\herm\Fbf  \end{array} \right].
\eeq
Now, for any $\lambda \in \C$, define the matrix
\beq \label{eq:Hlam}
    \Hbf_\lambda := \lambda\Ibf - \Hbf =
    \left[ \begin{array}{cc} \lambda\Ibf-\Dbf_s & -\Fbf \\
    \Fbf\herm\Dbf_s & \lambda\Ibf-\Dbf_x+\Fbf\herm\Fbf  \end{array} \right].
\eeq
For stability, we need to show that for any $|\lambda| \geq 1$, $\Hbf_\lambda$
is invertible.  We simplify this condition as follows:
Consider any $\lambda$ with $|\lambda| \geq 1$.
Now, $\Dbf_s$ in \eqref{eq:Dsdef} is a diagonal matrix with entries in $[0,1)$.
Hence $\lambda \Ibf -\Dbf_s$ is invertible since $|\lambda| \geq 1$.
Therefore, taking a Schur complement, we see that $\Hbf_\lambda$
is invertible if and only if the matrix
\beqan
    \Jbf_\lambda &:=& \lambda\Ibf-\Dbf_x+\Fbf\herm\Fbf +
        \Fbf\herm\Dbf_s(\lambda\Ibf-\Dbf_s)^{-1}\Fbf \nonumber \\
        &=& \lambda\Ibf-\Dbf_x+ \lambda\Fbf\herm(\lambda\Ibf-\Dbf_s)^{-1}\Fbf.
\eeqan
is invertible.  We can summarize the result as follows.

\begin{lemma}  \label{lem:Jlam}
Consider the GAMP Algorithm~\ref{algo:gamp} for any scalar estimation
functions satisfying the conditions in Section~\ref{sec:localStab}
including \eqref{eq:gderivbnd}.  The GAMP algorithm is locally stable
around a fixed point if and only if $\Jbf_\lambda$ is
invertible for all $|\lambda| \geq 1$, where
\beq \label{eq:Jdef}
    \Jbf_\lambda = \lambda\Ibf-\Dbf_x+ \lambda\Fbf\herm(\lambda\Ibf-\Dbf_s)^{-1}\Fbf,
\eeq
and $\Fbf$ is given in \eqref{eq:Fdef}.  In the special case of Gaussian
estimation functions \eqref{eq:gsxLin},
the above condition implies the GAMP Algorithm~\ref{algo:gamp},
will be globally stable.
\end{lemma}

\medskip
A similar calculation can be performed for the GAMP algorithm with
scalar stepsizes.  In this case, the vector stepsizes such as $\taubf_x$
and $\nubf_s$ are replaced with the scalar quantities $\tau_x$ and $\nu_s$.
For the case of Gaussian estimation functions \eqref{eq:gsxLin}
and identical variances \eqref{eq:constVar} we obtain the following:

\medskip
\begin{lemma}  \label{lem:JlamUnif}
Consider the GAMP Algorithm~\ref{algo:gampUnif} with
scalar stepsizes, Gaussian scalar estimation functions
\eqref{eq:gsxLin} and identical variances \eqref{eq:constVar}.
Then, the algorithm is globally
stable if and only if $\Jbf_\lambda$ is
invertible for all $|\lambda| \geq 1$, where
\beq \label{eq:JdefUnif}
    \Jbf_\lambda = (\lambda-d_x)\Ibf+
        \frac{\lambda}{\lambda-d_s}\Fbf\herm\Fbf,
\eeq
where
\beq \label{eq:FdefUnif}
    \Fbf = \sqrt{\theta_s\theta_x\nu_s\tau_x}\Abf,
\eeq
and
\begin{subequations} \label{eq:dsxUnif}
\beqa
    d_s &=& (1-\theta_s) + \theta_sq_s, \quad
        q_s = \frac{\nu_w}{\nu_p + \nu_w}, \\
    d_x &=& (1-\theta_x) + \theta_xq_x, \quad
        q_x = \frac{\tau_0}{\tau_0 + \tau_r}.
\eeqa
\end{subequations}
\end{lemma}

\section{Proof of Theorem \ref{thm:linConvUnif}} \label{sec:linConvUnifPf}

Our first step in the proof is to simplify the condition
in Lemma~\ref{lem:JlamUnif}.

\begin{lemma} \label{lem:HstabUnif}  Consider the GAMP
algorithm with scalar stepsizes, Algorithm~\ref{algo:gampUnif},
with the Gaussian scalar estimation functions \eqref{eq:gsxLin} and fixed
stepsizes.  Then the system is stable if and only if
\beq \label{eq:FcondUnif}
    \sigma^2_{\max}(\Abf)  < \|\Abf\|^2_F\gamma,
\eeq
where
\beq \label{eq:gamUnif1}
    \gamma := \frac{1}{\|\Abf\|^2_F\theta_s\theta_x}\left[ \frac{2}{\tau_x} - \frac{\theta_x}{\tau_0}
    \right]\left[  \frac{2}{\nu_s} - \frac{\theta_s}{\nu_w} \right].
\eeq
\end{lemma}
\begin{IEEEproof}  From Lemma~\ref{lem:JlamUnif},  we know that the system is
stable if and only if $\Jbf_\lambda$ in \eqref{eq:JdefUnif} is invertible
for all $|\lambda| \geq 1$.  To evaluate this condition,
suppose that $\Jbf_\lambda$ is not invertible for some $|\lambda| \geq 1$.
Then, there exists an $\vbf \neq 0$ such that $\Jbf_\lambda\vbf=0$,
which implies that
\[
    \Fbf\herm\Fbf\vbf = \frac{(d_x-\lambda)(\lambda-d_s)}{\lambda}\vbf.
\]
Using the expression for $\Fbf$ in \eqref{eq:FdefUnif}, this is equivalent to
\[
      \Abf\herm\Abf\vbf = \frac{(d_x-\lambda)(\lambda-d_s)}
      {\theta_x\theta_s\tau_x\nu_s\lambda}\vbf.
\]
Thus, $\vbf$ is an eigenvector of $\Abf\herm\Abf$.  But, $\sigma^2$ is an eigenvalue
of $\Abf\herm\Abf$ if and only if $\sigma$ is a singular value of $\Abf$.  Hence,
we conclude that $\Jbf_\lambda$ is non-invertible if and only if there exists a
singular value $\sigma$ of $\Abf$ such that
\[
    \sigma^2\theta_x\theta_z\tau_x\nu_s\lambda = (d_x-\lambda)(\lambda-d_s).
\]
Equivalently, we have shown that the system is stable if and only if the
the second-order polynomial
\[
    p(\lambda) := \lambda^2 +
    (\sigma^2\theta_x\theta_s\tau_x\nu_s-d_x-d_s)\lambda + d_sd_x
\]
has stable roots for all singular values of $\Abf$, $\sigma$.
Now recall that $d_s$ and $d_x \in (0,1)$.
By the Jury stability condition, the $p(\lambda)$ has stable roots if and only
$p(1) > 0$ and $p(-1) > 0$.  Now, the first condition is always satisfied since
\[
    p(1) = \sigma^2\theta_x\theta_z\tau_x\nu_s + (1-d_s)(1-d_x) > 0.
\]
So, the polynomial is stable if and only if
\[
    0 < p(-1) = -\sigma^2\theta_x\theta_z\tau_x\nu_s + (1+d_s)(1+d_x),
\]
or equivalently,
\[
    \sigma^2\theta_x\theta_z\tau_x\nu_s < (1+d_s)(1+d_x).
\]
For this to be true for all singular values of $\Abf$, we need
\[
     \sigma_{\max}^2(\Abf)\theta_x\theta_z\tau_x\nu_s < (1+d_s)(1+d_x).
\]
Thus, the system is stable if and only if \eqref{eq:FcondUnif}
is satisfied with
\beq \label{eq:gamUnif2}
   \gamma := \frac{(1+d_x)(1+d_s)}{\theta_s\theta_x\nu_s\tau_x\|\Abf\|^2_F}.
\eeq
So, we simply need to prove that \eqref{eq:gamUnif2} matches the
definition in \eqref{eq:gamUnif1}.  To this end, first note that
\beq \label{eq:dxfracUnif}
    \frac{1+d_x}{\tau_x} \stackrel{(a)}{=} \frac{2-\theta_x}{\tau_x}
        + \frac{\theta_x}{\tau_r}
    \stackrel{(b)}{=} \frac{2}{\tau_x} - \frac{\theta_x}{\tau_0},
\eeq
where (a) follows from the definition $q_x = \tau_x/\tau_r$ in
\eqref{eq:Dxdef} and (b) follows from the fixed-point equation \eqref{eq:tauxFixUnif}.
Similarly using \eqref{eq:Dsdef} and \eqref{eq:tausFixUnif},
 we obtain that
\beq  \label{eq:dsfracUnif}
    \frac{1+d_s}{\nu_s} = \frac{2-\theta_s}{\nu_s} + \frac{\theta_s}{\nu_p}
    = \frac{2}{\nu_s} - \frac{\theta_s}{\nu_w}.
\eeq
Substituting \eqref{eq:dxfracUnif} and \eqref{eq:dsfracUnif} into
\eqref{eq:gamUnif2}, we obtain \eqref{eq:gamUnif1} and the
lemma is proven.
\end{IEEEproof}

\medskip
\bluecolor
Let
\beq \label{eq:gamMin1}
    \Gamma := \inf_{\nu_w > 0} \gamma,
\eeq
where $\gamma$ is defined in \eqref{eq:gamUnif1} and the minimization is over $\nu_w$ with the
other parameters, $\|\Abf\|^2_F$, $\tau_0$, $m$ and $n$, being fixed.
It follows that if
\[
    \sigma^2(\Abf) < \Gamma\|\Abf\|^2_F
\]
then the system is stable for all $\nu_w$.
Conversely, if
\[
    \sigma^2(\Abf) > \Gamma\|\Abf\|^2_F
\]
then there exists at least one $\nu_w$ such that the system
is unstable.  So, the theorem will be proven if we can show that
$\Gamma$ defined in \eqref{eq:gamMin1} matches the expression
in \eqref{eq:gamUnif}.

To calculate the minima in \eqref{eq:gamMin1},
it is useful to write a scaled version of the updates.
Let
\begin{subequations} \label{eq:scale_def}
\begin{align}
  s & := \frac{m}{\|\Abf\|^2_F\nu_s\tau_0}, \quad
  x  := \frac{\tau_0}{\tau_x} \\
  u & :=  \frac{m}{\|\Abf\|^2_F\nu_w\tau_0}, \quad
  \beta := \frac{m}{n}.
\end{align}
\end{subequations}
Then, the fixed points of \eqref{eq:tauFixUnif} are given by
\beq \label{eq:sxFix}
  s = \frac{1}{x} + u, \quad x = \frac{\beta}{s} + 1.
\eeq
Also, $\gamma$ in \eqref{eq:gamUnif1} is given by,
\beq \label{eq:gamsx}
    \gamma = \frac{1}{m\theta_s\theta_x}(2x - \theta_x)(2s - \theta_su).
\eeq
Moreover, the minimization in \eqref{eq:gamMin1} is equivalent to
\beq \label{eq:gamMin2}
    \Gamma = \inf_{u \geq 0} \gamma,
\eeq
since minimizing over $\nu_w$ is equivalent to minimizing over $u$ in the scaled system.
To evaluate the minima \eqref{eq:gamMin2}, we first prove the following.

\begin{lemma}  The minimization in \eqref{eq:gamMin2} is given by
\beq \label{eq:gamMin3}
    \Gamma =  \lim_{u \arr 0} \gamma.
\eeq
That is, the minima is achieved as $u \arr 0$.
\end{lemma}
\begin{IEEEproof}
\textb{
From \eqref{eq:sxFix},
\beq \label{eq:sxFix1}
    \frac{u\beta}{s} = s-u+\beta-1.
\eeq
Substituting \eqref{eq:sxFix} into \eqref{eq:gamsx} and applying \eqref{eq:sxFix1}, we obtain
\begin{align}
     \gamma &= \frac{1}{m\theta_s\theta_x}\left(\frac{2\beta}{s} + 2 - \theta_x\right)(2s - \theta_su) \nonumber \\
    &= \frac{1}{m\theta_s\theta_x}\left[4\beta - (2-\theta_x)\theta_s u + 2(2-\theta_x)s - \frac{2\beta\theta_su}{s} \right] \nonumber \\
    &= \frac{1}{m\theta_s\theta_x}\left[ A(s,u) + B \right], \label{eq:gamsx2}
\end{align}
where
\begin{align} \label{eq:ABpf}
\begin{split}
    A(s,u) &:= 2(2-\theta_x-\theta_s)s + \theta_x\theta_s u \\
    B &:= 4\beta-2\theta_s(\beta-1)
\end{split}
\end{align}
Now let $s'$, $x'$ and $A'(s,u)$ denote the derivatives with respect to $u$.
From \eqref{eq:sxFix} we have
\beq
  s' = -\frac{x'}{x^2} + 1, \quad x' = -\frac{\beta s'}{s^2},
\eeq
and therefore,
\beq \label{eq:sDeriv}
    s' = \frac{s^2x^2}{s^2x^2-\beta}.
\eeq
Now from \eqref{eq:sxFix}, we have
\[
    sx > 1 \mbox{ and } sx > \beta.
\]
Therefore, $(sx)^2 > \beta$ and hence, from \eqref{eq:sDeriv}, $s' > 0$.
It follows that
\[
    A'(s,u) =  2(2-\theta_x-\theta_s)s' + \theta_x\theta_s > 0,
\]
since both $2-\theta_x-\theta_s \geq 0$ and $\theta_x\theta_x > 0$.
Hence, from \eqref{eq:gamsx2}, we have
\[
    \frac{\partial \gamma}{\partial u} = \frac{A'(s,u)}{m\theta_s\theta_x} > 0,
\]
and it follows that
the $\gamma$ is minimized by taking $u$ as small as possible.  Therefore,
\[
    \Gamma = \inf_{u \geq 0} \gamma = \lim_{u \arr 0} \gamma.
\]
}
\end{IEEEproof}
\color{black}

We conclude by evaluating the limit in \eqref{eq:gamMin3}.
The following lemma shows that value of the  minimization
agrees with \eqref{eq:gamUnif}, and hence completes the
proof of the theorem.

\bluecolor
\begin{lemma}  For any damping constants $\theta_s$, $\theta_x$,
the limit in \eqref{eq:gamMin3} is given by \eqref{eq:gamUnif}.
\end{lemma}
\begin{IEEEproof}  First consider the case when $\beta \geq 1$ (i.e.\ $m\geq n$).
In this case, as $u \arr 0$ the solutions to the fixed points \eqref{eq:sxFix}
will satisfy $s \arr 0$ and $x \arr \infty$.  Hence,
the limit of $A(s,u)$ in \eqref{eq:ABpf} is
\[
    \lim_{u \arr 0} A(s,u) = 0.
\]
Therefore,
\begin{align*}
    \Gamma &= \lim_{u \arr 0}  \gamma \stackrel{(a)}{=}  \frac{B}{m\theta_s\theta_x}
         \stackrel{(b)}{=} \frac{4\beta-2\theta_s(\beta-1)}{m\theta_s\theta_x} \\
         &\stackrel{(c)}{=} \frac{2\left[(2-\theta_s)m + \theta_sn\right]}{\theta_s\theta_x mn},
\end{align*}
where (a) used \eqref{eq:gamsx2}; (b) used \eqref{eq:ABpf} and (c) used the fact that $\beta=m/n$.
This proves the $m \geq n$ case of \eqref{eq:gamUnif}.

For the case when $\beta < 1$ (i.e.\ $m < n$)
and $u=0$, the solutions to fixed point in \eqref{eq:sxFix} are
\[
    x = \frac{1}{1-\beta}, \quad s = \frac{1}{x} = 1-\beta.
\]
Substituting $s=1-\beta$ and $u=0$ into \eqref{eq:gamsx2},
\begin{align*}
    \gamma &= \frac{1}{m\theta_s\theta_x}\left[ 2(2-\theta_x-\theta_s)(1-\beta)
    + 4\beta-2\theta_s(\beta-1) \right] \nonumber \\
    &= \frac{2\left[(2-\theta_x)n + \theta_xm\right]}{\theta_s\theta_x mn},
\end{align*}
where again we have used the fact that $\beta = m/n$.  Therefore,
\[
    \Gamma = \lim_{u \arr 0} \gamma  = \frac{2\left[(2-\theta_x)n + \theta_xm\right]}{\theta_s\theta_x mn},
\]
and this proves the $m < n$ case of \eqref{eq:gamUnif}.
\end{IEEEproof}
\color{black}

\section{Proof of Theorem~\ref{thm:localStab} } \label{sec:localStabPf}

We begin with a technical lemma.

\bluecolor
\begin{lemma} \label{lem:zbnd}
Let $\lambda \in \C$, $d_{s,\max}, d_x, \sigma \in [0,1)$ with $|\lambda|\geq 1$.  Define the set,
\beq \label{eq:Psetlam}
    P := \left\{ \lambda-d_x + \frac{\sigma^2\lambda}{\lambda - d_s} ~\mid
        ~d_s \in [0,d_{s,\max}] ~\right\}.
\eeq
Then $0 \not \in \mbox{conv}(P)$, the convex hull of $P$.
\end{lemma}

\begin{IEEEproof}
Write $\lambda$ in polar coordinates, $\lambda = re^{i\theta}$.
We first consider the case where $\theta \in (0,\pi)$.
Under this assumption, we claim for all $z \in P$,
\beq \label{eq:zimag}
    \Imag((\bar{\lambda} - d_x)z) < 0.
\eeq
Since $P$ is compact, this would imply
that \eqref{eq:zimag} holds for all $z \in \mbox{conv}(P)$.  In particular,
$0 \not \in \mbox{conv}(P)$.  So, we need to show that \eqref{eq:zimag} holds for
all $z \in P$.  

To this end, let $z \in P$ so that,
\beq \label{eq:zlem}
    z =  \lambda-d_x + \frac{\sigma^2\lambda}{\lambda - d_s},
\eeq
for some $d_s \in [0,d_{s,\max}]$.  Then,
\begin{align}
    \MoveEqLeft \Imag((\bar{\lambda} - d_x)z) \nonumber \\ 
    &= \Imag\left[|\lambda-d_x|^2 + \frac{\sigma^2(\bar{\lambda} - d_x)\lambda}{\lambda-d_s}\right] \nonumber \\
    &= \frac{\sigma^2}{|\lambda-d_s|^2} \Imag\left[(\bar{\lambda} - d_x)(\bar{\lambda} - d_s)\lambda\right] 
    \nonumber \\
    &= \frac{\sigma^2}{|\lambda-d_s|^2} \Imag\left[r^2\bar{\lambda} - (d_s+d_x)|\lambda|^2 + d_sd_x\lambda \right]
    \nonumber \\
    &= \frac{\sigma^2}{|\lambda-d_s|^2} \left[ -r^3\sin\theta + rd_sd_x\sin\theta \right]
    \nonumber  \\
    &= \frac{r\sin\theta \sigma^2}{|\lambda-d_s|^2} \left[ -r^2 + d_sd_x \right].    
    \label{eq:zimagsin}
\end{align}
Now, since $\theta \in (0,\pi)$, $\sin\theta > 0$.  Also, since $|\lambda|\geq 1$, $r \geq 1$.
Therefore, $r^2 > d_sd_x$ since $d_s,d_x < 1$.  Hence, \eqref{eq:zimagsin} shows that \eqref{eq:zimag}
holds for all $z \in P$.

Similarly, for the case when $\theta \in (-\pi,0)$, \eqref{eq:zimagsin} shows that
\beq \label{eq:zimag2}
    \Imag((\bar{\lambda} - d_x)z) > 0,
\eeq
for all $z \in P$.  The same argument then shows that $0 \not \in \mbox{conv}(P)$.

It remains to consider the cases when $\theta = 0$ or $\theta = \pi$.  For $\theta=0$,
$\lambda = r$ and any $z \in P$ is of the form,
\[
    z =  r-d_x + \frac{\sigma^2r}{r - d_s} \stackrel{(a)}{>} r-d_x \stackrel{(b)}{>} 0,
\]
where (a) follows from the fact that $r > d_s$ and (b) follows from the fact that $r>d_x$.
So, for all $z \in P$, $z$ is real and positive.  Hence, $0 \not \in \mbox{conv}(P)$.
Similarly, when $\theta = \pi$, $\lambda=-r$ and
\[
    z =  -r-d_x + \frac{\sigma^2r}{r + d_s} 
    < -r-d_x + \sigma^2 
    \stackrel{(a)}{<} -r + \sigma^2 \stackrel{(b)}{<} 0,
\]
where (a) follows since $d_x > 0$ and (b) follows since $r \geq 1$ and $\sigma^2 <1$.
Therefore, for all $z \in P$, $z$ is real and negative. Hence,
$0 \not \in \mbox{conv}(P)$.  We have thus shown that $0 \not \in \mbox{conv}(P)$
for all values of $\theta$.
\end{IEEEproof}
\color{black}

\medskip
We can now prove the main result.  Suppose that \eqref{eq:sigBnd} is satisfied.
By the definition of $\Fbf$ in \eqref{eq:Fdef} and $\tilde{\Abf}$ in
\eqref{eq:Anorm},  we have that
\beq \label{eq:FbndPf}
    \sigma^2_{\max}(\Fbf) < 1.
\eeq
Now, from Lemma~\ref{lem:Jlam} we need to show that
the matrix $\Jbf_\lambda$ in \eqref{eq:Jdef} is invertible
for all $\lambda \in \C$ with $|\lambda| \geq 1$.
We prove this by contradiction.

Suppose that $\Jbf_\lambda$ in \eqref{eq:Jdef} is not invertible for
some $\lambda$ with $|\lambda| \geq 1$.  Then, there exists an $\xbf$
with $\|\xbf\|^2=1$ such that $\xbf\herm\Jbf_\lambda\xbf = 0$.  Therefore,
if we define $\ybf = \Fbf\xbf$, the definition of $\Jbf_\lambda$ in \eqref{eq:Jdef}
shows that
\[
    \xbf\herm(\lambda\Ibf - \Dbf)\xbf + \lambda \ybf\herm(\lambda\Ibf-\Dbf_s)^{-1}\ybf = 0.
\]
Since $\Dbf_x$ and $\Dbf_s$ are diagonal, we have
\beq \label{eq:convsum}
    \sum_{j=1}^n (\lambda-d_{x_j}) |x_j|^2 +
    \sum_{i=1}^m \frac{\lambda}{\lambda-d_{s_j}}|y_j|^2 = 0.
\eeq
Since $\|\xbf\|^2=1$, we have $\sum_j |x_j|^2=1$.
Also, since $\|\Fbf\|_2^2= \sigma^2_{\max}(\Fbf)<1$,
\[
    \sum_i |y_i|^2 = \|\Fbf\xbf\|^2 = \sigma^2 \|\xbf\|^2 = \sigma^2
\]
for some $\sigma^2 < 1$.  Therefore,  \eqref{eq:convsum} shows that
\bluecolor
\beq \label{eq:zeroP}
    0 \in \mbox{conv}(P),
\eeq
where $P$ is the set \eqref{eq:Psetlam} where
\beq \label{eq:dsxlem}
    d_x = \sum_{j=1}^n d_{x_j} |x_j|^2, \quad d_{s,\max} = \max_{j} d_{s_j}.
\eeq
Now, from \eqref{eq:QsxDeriv} and the contractivity assumption \eqref{eq:gderivbnd},
the elements of the diagonal matrices $\Qbf_x$ and $\Qbf_s$ must be in the interval $(0,1)$.
Hence, from \eqref{eq:Dsxdef},
the elements $d_{x_j}$ and $d_{s_j} \in (0,1)$.  Therefore, 
 $d_x, d_{s,\max}$ in \eqref{eq:dsxlem} are in $(0,1)$.
From Lemma~\ref{lem:zbnd}, $0 \not \in \mbox{conv}(P_\lambda)$ which is a contradiction of \eqref{eq:zeroP}.
\color{black}
Hence, the assumption that $\Jbf_\lambda$
is not invertible must be false, and the theorem is proven.
}

\bibliographystyle{IEEEtran}
\bibliography{../bibl}

\newcommand{\SortNoop}[1]{}
\begin{thebibliography}{10}
\providecommand{\url}[1]{#1}
\csname url@samestyle\endcsname
\providecommand{\newblock}{\relax}
\providecommand{\bibinfo}[2]{#2}
\providecommand{\BIBentrySTDinterwordspacing}{\spaceskip=0pt\relax}
\providecommand{\BIBentryALTinterwordstretchfactor}{4}
\providecommand{\BIBentryALTinterwordspacing}{\spaceskip=\fontdimen2\font plus
\BIBentryALTinterwordstretchfactor\fontdimen3\font minus
  \fontdimen4\font\relax}
\providecommand{\BIBforeignlanguage}[2]{{%
\expandafter\ifx\csname l@#1\endcsname\relax
\typeout{** WARNING: IEEEtran.bst: No hyphenation pattern has been}%
\typeout{** loaded for the language `#1'. Using the pattern for}%
\typeout{** the default language instead.}%
\else
\language=\csname l@#1\endcsname
\fi
#2}}
\providecommand{\BIBdecl}{\relax}
\BIBdecl

\bibitem{RanSchFle:14-ISIT}
S.~Rangan, P.~Schniter, and A.~K. Fletcher, ``On the convergence of approximate
  message passing with arbitrary matrices,'' in \emph{Proc. IEEE ISIT}, Jul.
  2014, pp. 236--240.

\bibitem{ChamDLL:98}
A.~Chambolle, R.~A. DeVore, N.~Y. Lee, and B.~J. Lucier, ``Nonlinear wavelet
  image processing: Variational problems, compression, and noise removal
  through wavelet shrinkage,'' \emph{IEEE Trans. Image Process.}, vol.~7,
  no.~3, pp. 319--335, Mar. 1998.

\bibitem{DaubechiesDM:04}
I.~Daubechies, M.~Defrise, and C.~D. Mol, ``An iterative thresholding algorithm
  for linear inverse problems with a sparsity constraint,'' \emph{Commun. Pure
  Appl. Math.}, vol.~57, no.~11, pp. 1413--1457, Nov. 2004.

\bibitem{WrightNF:09}
S.~J. Wright, R.~D. Nowak, and M.~Figueiredo, ``Sparse reconstruction by
  separable approximation,'' \emph{IEEE Trans. Signal Process.}, vol.~57,
  no.~7, pp. 2479--2493, Jul. 2009.

\bibitem{BeckTeb:09}
A.~Beck and M.~Teboulle, ``A fast iterative shrinkage-thresholding algorithm
  for linear inverse problem,'' \emph{SIAM J.\ Imag.\ Sci.}, vol.~2, no.~1, pp.
  183–--202, 2009.

\bibitem{Nesterov:07}
Y.~E. Nesterov, ``Gradient methods for minimizing composite objective
  function,'' center for Operations Research and Econometrics (CORE), Catholic
  Univ. Louvain, Louvain-la-Neuve, Belgium, CORE Discussion Paper 2007/76,
  2007.

\bibitem{BioDFig:07}
J.~Bioucas-Dias and M.~Figueiredo, ``A new {TwIST}: Two-step iterative
  shrinkage/thresholding algorithms for image restoration,'' \emph{IEEE Trans.
  Image Process.}, vol.~16, no.~12, pp. 2992 -- 3004, Dec. 2007.

\bibitem{BoydPCPE:09}
S.~Boyd, N.~Parikh, E.~Chu, B.~Peleato, and J.~Eckstein, ``Distributed
  optimization and statistical learning via the alternating direction method of
  multipliers,'' \emph{Found. Trends Mach. Learn.}, vol.~3, pp. 1--122, 2010.

\bibitem{Esser:JIS:10}
E.~Esser, X.~Zhang, and T.~F. Chan, ``A general framework for a class of first
  order primal-dual algorithms for convex optimization in imaging science,''
  \emph{SIAM J. Imaging Sci.}, vol.~3, no.~4, pp. 1015--1046, 2010.

\bibitem{Chambolle:JMIV:11}
A.~Chambolle and T.~Pock, ``A first-order primal-dual algorithm for convex
  problems with applications to imaging,'' \emph{J. Math. Imaging Vis.},
  vol.~40, pp. 120--145, 2011.

\bibitem{He:JIS:12}
B.~He and X.~Yuan, ``Convergence analysis of primal-dual algorithms for a
  saddle-point problem: From contraction perspective,'' \emph{SIAM J. Imaging
  Sci.}, vol.~5, no.~1, pp. 119--149, 2012.

\bibitem{Komodakis:SPM:15}
N.~Komodakis and J.-C. Pesquet, ``Playing with duality: {A}n overview of recent
  primal-dual approaches for solving large-scale optimization problems,''
  \emph{IEEE Signal Process. Mag.}, vol.~32, no.~6, pp. 31--54, 2015.

\bibitem{DonohoMM:09}
D.~L. Donoho, A.~Maleki, and A.~Montanari, ``Message-passing algorithms for
  compressed sensing,'' \emph{Proc. Nat. Acad. Sci.}, vol. 106, no.~45, pp.
  18\,914--18\,919, Nov. 2009.

\bibitem{DonohoMM:10-ITW1}
------, ``Message passing algorithms for compressed sensing {I}: {M}otivation
  and construction,'' in \emph{Proc.\ Info.\ Theory Workshop}, Jan. 2010, pp.
  1--5.

\bibitem{Rangan:11-ISIT}
S.~Rangan, ``Generalized approximate message passing for estimation with random
  linear mixing,'' in \emph{Proc. IEEE ISIT}, 2011, pp. 2174--2178.

\bibitem{BayatiM:11}
M.~Bayati and A.~Montanari, ``The dynamics of message passing on dense graphs,
  with applications to compressed sensing,'' \emph{IEEE Trans. Inform. Theory},
  vol.~57, no.~2, pp. 764--785, Feb. 2011.

\bibitem{javanmard2013state}
A.~Javanmard and A.~Montanari, ``State evolution for general approximate
  message passing algorithms, with applications to spatial coupling,''
  \emph{Information and Inference}, vol.~2, no.~2, pp. 115--144, 2013.

\bibitem{BayLelMon:15}
M.~Bayati, M.~Lelarge, and A.~Montanari, ``Universality in polytope phase
  transitions and message passing algorithms,'' \emph{Ann. Appl. Prob.},
  vol.~25, no.~2, pp. 753--822, 2015.

\bibitem{rush2016finite}
C.~Rush and R.~Venkataramanan, ``Finite-sample analysis of approximate message
  passing,'' in \emph{Proc.\ IEEE ISIT}, 2016, pp. 755--759.

\bibitem{RanSRFC:13-ISIT}
S.~Rangan, P.~Schniter, E.~Riegler, A.~Fletcher, and V.~Cevher, ``Fixed points
  of generalized approximate message passing with arbitrary matrices,'' in
  \emph{Proc. IEEE ISIT}, Jul. 2013, pp. 664--668.

\bibitem{Krzakala:14-ISITbethe}
F.~Krzakala, A.~Manoel, E.~W. Tramel, and L.~Zdeborov{\'a}, ``Variational free
  energies for compressed sensing,'' in \emph{Proc. IEEE ISIT}, Jul. 2014, pp.
  1499--1503.

\bibitem{YedidiaFW:03}
J.~S. Yedidia, W.~T. Freeman, and Y.~Weiss, ``Understanding belief propagation
  and its generalizations,'' in \emph{Exploring Artificial Intelligence in the
  New Millennium}.\hskip 1em plus 0.5em minus 0.4em\relax San Francisco, CA:
  Morgan Kaufmann Publishers, 2003, pp. 239--269.

\bibitem{Vila:ICASSP:15}
J.~Vila, P.~Schniter, S.~Rangan, F.~Krzakala, and L.~Zdeborov{\'a}, ``Adaptive
  damping and mean removal for the generalized approximate message passing
  algorithm,'' in \emph{Proc. IEEE ICASSP}, 2015, pp. 2021--2025.

\bibitem{Caltagirone:14-ISIT}
F.~Caltagirone, L.~Zdeborov{\'a}, and F.~Krzakala, ``On convergence of
  approximate message passing,'' in \emph{Proc. IEEE ISIT}, Jul. 2014, pp.
  1812--1816.

\bibitem{pretti2005message}
M.~Pretti, ``A message-passing algorithm with damping,'' \emph{Journal of
  Statistical Mechanics: Theory and Experiment}, vol. 2005, no.~11, p. P11008,
  2005.

\bibitem{kolmogorov2006convergent}
V.~Kolmogorov, ``Convergent tree-reweighted message passing for energy
  minimization,'' \emph{IEEE Trans.\ Pattern Analysis and Machine
  Intelligence}, vol.~28, no.~10, pp. 1568--1583, 2006.

\bibitem{globerson2007fixing}
A.~Globerson and T.~S. Jaakkola, ``Fixing max-product: Convergent message
  passing algorithms for map lp-relaxations,'' in \emph{Proc.\ NIPS}, 2007, pp.
  553--560.

\bibitem{manoel2015swamp}
A.~Manoel, F.~Krzakala, E.~W. Tramel, and L.~Zdeborov{\'a}, ``Swept approximate
  message passing for sparse estimation,'' in \emph{Proc. ICML}, 2015, pp.
  1123--1132.

\bibitem{rangan2015admm}
S.~Rangan, A.~K. Fletcher, P.~Schniter, and U.~S. Kamilov, ``Inference for
  generalized linear models via alternating directions and {B}ethe free energy
  minimization,'' in \emph{Proc. IEEE ISIT}, 2015, pp. 1640--1644.

\bibitem{bickson2008gaussian}
D.~Bickson, ``Gaussian belief propagation: Theory and application,''
  \emph{arXiv:0811.2518}, 2008.

\bibitem{dolev2009fixing}
D.~Dolev, D.~Bickson, and J.~K. Johnson, ``Fixing convergence of {G}aussian
  belief propagation,'' in \emph{Proc.\ IEEE ISIT}, 2009, pp. 1674--1678.

\bibitem{Schniter:ALL:12}
P.~Schniter and S.~Rangan, ``Compressive phase retrieval via generalized
  approximate message passing,'' in \emph{Proc. Allerton Conf. Comm. Control \&
  Comput.}, Monticello, IL, Oct. 2012.

\bibitem{Rangan:10arXiv-GAMP}
S.~Rangan, ``Generalized approximate message passing for estimation with random
  linear mixing,'' arXiv:1010.5141v1 [cs.IT]., Oct. 2010.

\bibitem{Combettes:MMS:05}
P.~L. Combettes and V.~R. Wajs, ``Signal recovery by proximal forward-backward
  splitting,'' \emph{Multiscale Model. Simul.}, vol.~4, pp. 1168--1200, 2005.

\bibitem{Arrow:Book:58}
K.~J. Arrow, L.~Hurwicz, and H.~Uzawa, \emph{Studies in Linear and Non-Linear
  Programming}.\hskip 1em plus 0.5em minus 0.4em\relax Palo Alto, CA: Stanford
  University Press, 1958.

\bibitem{Goldstein:13}
T.~Goldstein, E.~Esser, and R.~Baraniuk, ``Adaptive primal-dual hybrid gradient
  methods for saddle-point problems,'' \emph{arXiv:1305.0546}, 2013.

\bibitem{MarcenkoP:67}
V.~A. Mar\v{c}enko and L.~A. Pastur, ``Distribution of eigenvalues for some
  sets of random matrices,'' \emph{Math. USSR--Sbornik}, vol.~1, no.~4, pp.
  457--483, 1967.

\bibitem{malioutov2006walk}
D.~M. Malioutov, J.~K. Johnson, and A.~S. Willsky, ``Walk-sums and belief
  propagation in {G}aussian graphical models,'' \emph{The Journal of Machine
  Learning Research}, vol.~7, pp. 2031--2064, 2006.

\bibitem{weiss2000correctness}
Y.~Weiss and W.~T. Freeman, ``Correctness of belief propagation in {G}aussian
  graphical models of arbitrary topology,'' in \emph{Advances in neural
  information processing systems}, 2000, pp. 673--679.

\bibitem{rusmevichientong2001analysis}
P.~Rusmevichientong and B.~Van~Roy, ``An analysis of belief propagation on the
  turbo decoding graph with {G}aussian densities,'' \emph{IEEE Trans. Inform.
  Theory}, vol.~47, no.~2, pp. 745--765, 2001.

\bibitem{moallemi2009convergence}
C.~C. Moallemi and B.~Van~Roy, ``Convergence of min-sum message passing for
  quadratic optimization,'' \emph{IEEE Trans. Inform. Theory}, vol.~55, no.~5,
  pp. 2413--2423, 2009.

\bibitem{MalioutovJW:06}
D.~M. Malioutov, J.~K. Johnson, and A.~S. Willsky, ``Walk-sums and belief
  propagation in {G}aussian graphical models,'' \emph{J. Machine Learning
  Res.}, vol.~7, pp. 2031--2064, Oct. 2006.

\bibitem{moallemi2010convergence}
C.~C. Moallemi and B.~Van~Roy, ``Convergence of min-sum message-passing for
  convex optimization,'' \emph{IEEE Trans. Inform. Theory}, vol.~56, no.~4, pp.
  2041--2050, 2010.

\bibitem{Vidyasagar:78}
M.~Vidyasagar, \emph{Nonlinear Systems Analysis}.\hskip 1em plus 0.5em minus
  0.4em\relax Englewood Cliffs, NJ: Prentice-Hall, 1978.

\bibitem{rangan2017vamp}
S.~Rangan, P.~Schniter, and A.~K. Fletcher, ``Vector approximate message
  passing,'' in \emph{Proc.\ IEEE ISIT}, 2017, pp. 1588--1592.

\bibitem{fletcher2017learning}
A.~K. Fletcher and P.~Schniter, ``Learning and free energies for vector
  approximate message passing,'' in \emph{IEEE Intl. Conf. Acoustics, Speech
  and Signal Processing (ICASSP)}, 2017, pp. 4247--4251.

\bibitem{yates:95}
R.~D. Yates, ``A framework for uplink power control in cellular radio
  systems,'' \emph{IEEE J. Sel. Areas Comm.}, vol.~13, no.~7, pp. 1341--1347,
  September 1995.

\end{thebibliography}
\end{document}